%% file: template.tex
\documentclass{sfuthesis}

\title{Comparative Analysis of Transfer Learning in Deep Learning Text-to-Speech Models on a Few-Shot, Low-Resource, Customized Dataset}
\thesistype{Thesis}
\author{Ze Liu}

\degree{Honours Degree of Bachelor of Science}
\department{Department of Computer Science}
\copyrightyear{2023}

\usepackage{amsmath}                            % (1)
\usepackage{amssymb}                            % (1)
\usepackage{amsthm}                             % (2)
\usepackage{graphicx}                           % (3)
\usepackage[
pdfborder={0 0 0}]{hyperref}        % (4)
% ...
% ...
% ...
\usepackage[utf8]{inputenc}
\usepackage[T1]{fontenc}
\usepackage{tipa}
\usepackage{svg}
\usepackage{tabularx}
\usepackage{apacite}
\usepackage[labelsep=newline,labelfont=bf]{caption}
\usepackage[utf8]{inputenc}
\usepackage{tabularx}
\usepackage{booktabs}

\usepackage{caption}
\renewcommand{\defaultspacing}{\doublespacing}  % (3)
% ...
% ...
% ...
\begin{document}

\frontmatter
\maketitle{}
% \makecommittee{}

%\addtoToC{Ethics Statement}%
%\includepdf[pagecommand={\thispagestyle{plain}}]{ethics_statement_piii.pdf}%
%\clearpage

\newpage % start a new page

\begin{center}
    % \vspace*{\fill} % center vertically
    A dissertation submitted to the Department of Computer Science\\
    in partial fulfillment of the requirements for the degree of\\
    Bachelor of Science (Honours) in Computer Science\\

    \vspace*{1in}
    August 2023
    % \vspace*{\fill} % center vertically
\end{center}

\begin{abstract}
Text-to-Speech (TTS) synthesis using deep learning relies on voice quality. Modern TTS models are advanced, but they need large amount of data. Given the growing computational complexity of these models and the scarcity of large, high-quality datasets, this research focuses on transfer learning, especially on few-shot, low-resource, and customized datasets. In this research, "low-resource" specifically refers to situations where there are limited amounts of training data, such as a small number of audio recordings and corresponding transcriptions for a particular language or dialect.

This thesis, is rooted in the pressing need to find TTS models that require less training time, fewer data samples, yet yield high-quality voice output. The research evaluates TTS state-of-the-art model transfer learning capabilities through a thorough technical analysis. It then conducts a hands-on experimental analysis to compare models' performance in a constrained dataset.
This study investigates the efficacy of modern TTS systems with transfer learning on specialized datasets and a model that balances training efficiency and synthesis quality. Initial hypotheses suggest that transfer learning could significantly improve TTS models' performance on compact datasets, and an optimal model may exist for such unique conditions. This thesis predicts a rise in transfer learning in TTS as data scarcity increases. In the future, custom TTS applications will favour models optimized for specific datasets over generic, data-intensive ones.
\end{abstract}

\begin{acknowledgements}
I would like to express my deepest appreciation to my supervisor, Dr. Xianta Jiang, whose invaluable guidance, support, and encouragement have been instrumental in the completion of this thesis. His profound knowledge, insightful feedback, and unwavering patience have been a source of constant inspiration, pushing me to explore beyond my limits and fostering a deep sense of curiosity and passion for the field of Computer Science.
\end{acknowledgements}

% 目录: Table of Contents

\addtoToC{Table of Contents}%
\tableofcontents%
\clearpage

%   This is an optional page. Remove the following lines if you don't have any tables.
\addtoToC{List of Tables}%
\listoftables%
\clearpage

%   This is an optional page. Remove the following lines if you don't have any figures.
\addtoToC{List of Figures}%
\listoffigures%
\clearpage

%   MAIN MATTER  %%%%%%%%%%%%%%%%%%%%%%%%%%%%%%%%%%%%%%%%%%%%%%%%%%%%%%%%%%%%%%
%
%   Start writing your thesis --- or start \include ing chapters --- here.
%

\mainmatter%

\chapter{Introduction}

In the vast landscape of deep learning, one of the critical challenges faced is the need for vast amounts of data to train effective models. This is especially pertinent in areas like Text-to-Speech (TTS) where the quality of synthesized speech is paramount. Given the ever-increasing computational complexity of modern TTS models and the ever-present challenge of gathering large, quality datasets, it becomes crucial to investigate the efficacy of transfer learning, especially on smaller, customized datasets. This becomes all the more significant in the context of applications that require rapid model deployment with high-quality output, using limited resources.

\section{Context of the Discipline}

Deep Learning-based TTS systems have set the benchmark for voice synthesis, shifting from the traditional concatenative and formant-based methods to neural architectures. These systems, while groundbreaking in their capabilities, are notorious for their hunger for data. As a result, the challenge arises when we deal with few-shot, low-resource, or customized datasets. Can these models, trained on vast general datasets, effectively adapt to smaller, more specialized datasets? Transfer learning promises a solution, allowing models to leverage knowledge acquired from one task to another, potentially offering a shortcut to high-quality synthesis without the need for exhaustive training.

\section{Objective of the Thesis}

Against this backdrop, this thesis titled, \textit{"Comparative Analysis of Transfer Learning in Deep Learning Text-to-Speech Models on a Few-Shot, Low-Resource, Customized Dataset,"} embarks on a journey to:

\begin{enumerate}
    \item \textbf{Technical Examination}: Systematically analyze leading TTS systems in the context of transfer learning capabilities.
    \item \textbf{Experimental Analysis}: Perform hands-on experiments to evaluate the performance of various models on a few-shot, low-resource, customized dataset.
    \item \textbf{Model Identification}: Identify a TTS model that strikes the optimal balance between training speed, minimal data requirement, and high-quality voice output.
\end{enumerate}

\noindent \textbf{Research Questions}:

\begin{itemize}
    \item How effective are modern TTS systems when employing transfer learning on small, customized datasets?
    \item Which model, among the contenders, offers the best trade-off between training speed, dataset size requirement, and synthesis quality?
\end{itemize}

\noindent \textbf{Hypotheses}:

\begin{itemize}
    \item Transfer learning significantly enhances the capability of TTS models to perform effectively on smaller datasets.
    \item There exists an optimal TTS model that offers a harmonious blend of rapid training, minimal data dependence, and exemplary voice quality.
\end{itemize}

\noindent \textbf{Predictions}:

\begin{itemize}
    \item Transfer learning will gain more traction in the TTS domain as data scarcity becomes more prominent.
    \item Customized TTS applications will lean towards models optimized for performance on specialized datasets rather than generic, data-hungry models.
\end{itemize}

As we delve deeper into the subsequent chapters, this thesis will unveil the findings from the comparative analysis, providing insights and guidance for future endeavors in the world of TTS, especially in constrained data scenarios.

% LITERATURE REVIEW  %%%%%%%%%%%%%%%%%%%%%%%%%%%%%%%%%%%%%%%%%%%%%%%%%%%%%%%%%%%%%%

% \chapter{Literature Review}
% \include{literature_review}
\include{literature_reivew_1}

 %%%%%%%%%%%%%%%%%%%%%%%%%%%%%%%%%%%%%%%%%%%%%%%%%%%%%%%%%%%%%%

% Methodology PART  %%%%%%%%%%%%%%%%%%%%%%%%%%%%%%%%%%%%%%%%%%%%%%%%%%%%%%%%%%%%%%

% \chapter{Literature Review}
\include{Methodology}

 %%%%%%%%%%%%%%%%%%%%%%%%%%%%%%%%%%%%%%%%%%%%%%%%%%%%%%%%%%%%%%

\include{experiment}

\include{Result}

\include{Discussion}

%   BACK MATTER  %%%%%%%%%%%%%%%%%%%%%%%%%%%%%%%%%%%%%%%%%%%%%%%%%%%%%%%%%%%%%%
%
%   References and appendices. Appendices come after the bibliography and
%   should be in the order that they are referred to in the text.
%
%   If you include figures, etc. in an appendix, be sure to use
%
%       \caption[]{...}
%
%   to make sure they are not listed in the List of Figures.
%

% \backmatter%
%     \addtoToC{Bibliography}
%     % \bibliographystyle{plain}
%     % \bibliography{references}
%     \addbibresource{references.bib}
% % 

\backmatter%
    \addtoToC{Bibliography}
    \bibliographystyle{apacite}
    \bibliography{references}

% \begin{appendices} % optional
% \chapter{Code}

% Appendices should be used for supplemental information that does not form part of the main research. Remember that figures and tables in appendices should not be listed in the List of Figures or List of Tables.

% \end{appendices}

\end{document}

%% file: literature_reivew_1.tex
\chapter{Literature Review}

\section{Early TTS Systems}
\subsection{Rule-Based Systems}
During the 1980s, rule-based systems were the prevailing approach in the field of Text-to-Speech (TTS), with a strong reliance on linguistic rules as shown by \cite{klatt1980software}. Phonemic representations of textual content were generated and then used as input for algorithms in order to generate corresponding auditory output, whereby prosodic attributes such as intonation and stress were manually calibrated.

\subsection{Concatenative Synthesis}
During the 1990s, concatenative synthesis gained prominence as the preferred approach \cite{hunt1996unit}. The continuous speech was generated by assembling and storing pre-recorded fragments of human voice. The system's ability to generate outputs that seem more realistic was limited due to the limited quantity and variety of the database.

\subsection{Statistical Models}
During the early 2000s, there was a surge in the popularity of Hidden Markov Model (HMM)-Based Synthesis, as shown by the work of \cite{tokuda2000speech}. Rather of retaining whole speech fragments, statistical models were used to capture speech characteristics. The aforementioned methodology exhibited adaptability and a reduced memory footprint but sometimes led to the manifestation of "muffled" or "flattened" vocal outputs \cite{tokuda2000speech}.

\section{Deep Learning Models}
\subsubsection{WaveNet}
WaveNet by DeepMind \cite{wavenet} improved voice synthesis, making it more realistic. It is raw audio waveforms instead of high-level properties or parameters distinguish WaveNet. It provides precise audio outputs with this fine-grained approach.
WaveNet's conditioning is another trait. This allows the model to generate sounds in multiple languages or impersonate speakers based on input conditions, boosting its application potential.
TTS systems produced mechanical audio before WaveNet. WaveNet made speech more natural. It reduced problematic audio splicing in concatenative TTS systems.
However, WaveNet had challenges. The model was computationally expensive, especially during generation, hence real-time applications needed robust hardware \cite{wavenet}. To perform successfully, early WaveNet needed lots of high-quality training data. The sample-by-sample audio generation delay was another issue. The model's various parameters required a lot of computing power to train \cite{wavenet}. WaveNet audio sometimes differs from training data.
WaveNet's speech synthesis achievement was noteworthy despite these challenges. It spurred research and technology to increase its limits while preserving audio quality.

\subsubsection{Parallel WaveNet}
To solve the problem of WaveNet being too resource intensive, deepMind's Parallel WaveNet \cite{2018parallel} and WaveGlow \cite{2019waveglow} models are improved models. WaveNet's quick and simultaneous audio generation makes it more efficient for real-time applications. The original WaveNet produced audio sample-by-sample, rendering it unsuitable for real-time synthesis. To solve this problem, Parallel WaveNet used "Inverse Autoregressive Flow" to create all audio samples simultaneously. Parallel WaveNet increased speed and maintained WaveNet's high-quality audio. This was a breakthrough since it allowed faster generation without sacrificing sound quality \cite{2018parallel}. This technology enabled real-time voice applications like digital assistants and interactive response systems \cite{2018parallel}. Parallel WaveNet's success shows that deep learning in speech synthesis may overcome the constraints and challenges of cutting-edge models with refinements and technological improvements \cite{2018parallel}.

\subsubsection{WaveGlow}
WaveGlow \cite{2019waveglow} is a flow-based generative neural network for speech synthesis that replaces WaveNet and Parallel WaveNet. NVIDIA's WaveGlow combines WaveNet's greatest features with the Gaussian mixture model-based Parallel WaveNet for quicker, higher-quality speech synthesis \cite{2019waveglow}. The main benefit of WaveGlow is its simplicity and efficiency. Its flow-based technique offers parallelism and quick audio synthesis while providing WaveNet-like audio quality. This means it can synthesize high-quality audio in real-time without the computational bottlenecks found in some models \cite{2019waveglow}. WaveGlow combines autoregressive models and normalizing flows to create a single model without probability density distillation or auxiliary models. This simplifies audio generation and training \cite{2019waveglow}.

\subsubsection{Tacotron}
Tacotron \cite{wang2017tacotron} was developed by Google. Prior to Tacotron, TTS systems had segmented models, each addressing specific synthesis stages. These often produced speech lacking in human-like prosody \cite{wang2017tacotron}. Tacotron, developed by Google's research team, aimed to streamline this, targeting natural intonation, and simplifying the intricate TTS pipeline \cite{wang2017tacotron}. Tacotron employed an end-to-end sequence-to-sequence model, reminiscent of techniques in machine translation. Central to its operation was an attention mechanism, enabling dynamic focus on the input text, thus efficiently learning text-audio alignment \cite{wang2017tacotron}. 
It provides numerous valuable contributions:
\begin{itemize}
    \item End-to-End Learning: Tacotron pioneered in demonstrating direct text-to-speech mapping, eliminating the need for hand-crafted features \cite{wang2017tacotron}.
    \item Enhanced Naturalness: With Tacotron, speech synthesis attained a new standard, rivaling contemporary TTS systems. This was amplified when combined with the WaveNet vocoder \cite{wang2017tacotron}.
    \item Streamlined TTS Pipeline: Tacotron's approach drastically reduced TTS complexities \cite{wang2017tacotron}.
\end{itemize}

However, Tacotron was not without its challenges. The attention mechanism, while innovative, occasionally introduced speech artifacts such as word repetitions \cite{shen2018natural}. Training the model was another intricate process; stability was an issue that often necessitated meticulous hyperparameter tuning \cite{shen2018natural}. Additionally, achieving real-time synthesis, particularly when integrating with WaveNet, presented significant obstacles. Furthermore, Tacotron had a voracious appetite for vast quantities of high-quality, annotated training data, and it occasionally faltered when confronted with rare words or unconventional pronunciations \cite{shen2018natural}.

\subsubsection{Tacotron2}

Tacotron 2 \cite{shen2018natural}, a successor to the original Tacotron, brought forth a series of refinements and innovations in the realm of text-to-speech synthesis  \cite{shen2018natural}. One of its standout enhancements was in its structural design. The encoder in Tacotron 2 integrated a higher number of convolutional layers  \cite{shen2018natural}, allowing it to adeptly capture text features, particularly beneficial for elongated text sequences  \cite{shen2018natural}. Additionally, by incorporating WaveNet as its vocoder, the produced speech was significantly more natural and lifelike  \cite{shen2018natural}.

In terms of attention mechanisms, while the original Tacotron had pioneered the concept, Tacotron 2 advanced this feature, ensuring more stable outcomes for extended texts  \cite{shen2018natural}. This stability was further boosted by Tacotron 2's simplified symbolic representation for input text, fostering a better text-to-speech mapping  \cite{shen2018natural}. Not stopping at that, Tacotron 2 introduced a post-processing network to refine the generated Mel spectrogram, which in turn elevated the overall speech quality  \cite{shen2018natural}.

Yet, as with many sophisticated models, Tacotron 2 had its set of challenges. Training stability, especially concerning the attention mechanism, often posed difficulties, more so with extended textual content. And while the integration of WaveNet augmented speech quality, it rendered real-time synthesis quite challenging due to WaveNet's slower generation pace  \cite{shen2018natural}. Achieving optimal performance with Tacotron 2 often mandated vast quantities of high-caliber, annotated training data \cite{shen2018natural}. Furthermore, the model sometimes grappled with rare words or unconventional pronunciations. Lastly, occasional inconsistencies in pronunciation or rhythm were observed in the generated speech \cite{shen2018natural}.

\subsubsection{Transformer TTS}
To address Conventional RNN and LSTM architectures have difficulties in processing long sequences, leading to inefficient training and long computational delays \cite{transformer}. Transformer TTS \cite{transformer} have taken inspiration from the achievements of Transformer designs in natural language processing (NLP) problems and subsequently used them to text-to-speech (TTS) systems. The design in question intrinsically incorporates self-attention processes, which have the potential to provide more stable and accurate alignment compared to the LSTM-based attention mechanism utilized in Tacotron \cite{transformer}.

\subsubsection{FastSpeech}
To address the speed and stability problems of Tacotron 2, FastSpeech \cite{ren2019fastspeech} was proposed. It uses a non-autoregressive approach and incorporates the structure of the Transformer.

The primary advancements of FastSpeech lie in its speed and determinism. Its non-autoregressive design enables parallelized prediction, hence improving both speed and determinism \cite{ren2019fastspeech}. 
The robustness of FastSpeech is improved by removing the attention mechanism, which mitigates frequent problems such as the occurrence of missed or repeated audio segments \cite{ren2019fastspeech}.
Monotonic Alignment is achieved through the implementation of a teacher-student training strategy in FastSpeech \cite{ren2019fastspeech}. This approach guarantees a consistent pronunciation and rhythm, resulting in a stable alignment between the given text and the corresponding speech output \cite{ren2019fastspeech}. 

While the FastSpeech system is considered groundbreaking, it also encounters several challenges and limitations. The training process of dependency parsing is further complicated by its need on a pre-existing model for aligning the data \cite{ren2020fastspeech}. Certain critiques raise the possibility of a trade-off in terms of expressiveness when comparing to attention-based alternatives \cite{ren2020fastspeech}.The implementation of the unique design presents challenges, since it provides a distinct set of difficulties while also addressing some issues \cite{ren2020fastspeech}.

% FastSpeech 2 (2020): 是FastSpeech的改进版本，它引入了更多的特性，如持续时间预测和声学特性回归，以进一步提高语音的自然度和质量。

\subsubsection{Parallel WaveGAN}
The speed of traditional vocoders, such as WaveNet, is quite slow, particularly for real-time synthesis, mostly because of their autoregressive nature, despite their ability to generate speech of high quality \cite{yamamoto2020parallel}.

In order to facilitate the integration of waveform generating methods into real-world applications, particularly on devices with constrained processing capabilities, there arose a necessity for enhanced computational efficiency and expedited performance \cite{yamamoto2020parallel}.

In contrast to WaveNet, which use a sequential approach by predicting individual samples, Parallel WaveGAN utilizes a non-autoregressive method. This implies that it produces many waveform samples simultaneously, resulting in a significant enhancement of the synthesis speed  \cite{yamamoto2020parallel}.

The structure of the Generative Adversarial Network (GAN) is characterized by parallel components. The WaveGAN model utilizes a Generative Adversarial Network (GAN) architecture \cite{yamamoto2020parallel}. The GAN configuration comprises a generator and a discriminator that engage in a game-theoretic competition. The primary objective of the generator is to generate waveforms that closely resemble genuine waveforms \cite{yamamoto2020parallel}, whereas the discriminator's primary goal is to accurately distinguish between real waveforms and those made by the generator. Over the course of its development, the generator gradually acquires the ability to generate waveforms of exceptional quality  \cite{yamamoto2020parallel}.

In order to uphold the quality of audio, the Parallel WaveGAN model incorporates a multi-resolution Short-Time Fourier Transform (STFT) loss  \cite{yamamoto2020parallel}. The preservation of both local and global structures of the audio waveform through this loss contributes to the production of synthetic speech of superior quality  \cite{yamamoto2020parallel}.

\subsubsection{FastSpeech2}
FastSpeech 2 was developed to address issues left unanswered by its predecessor. Due to its non-autoregressive and deterministic nature, FastSpeech had limited prosody and expressiveness. Despite its benefits, this deterministic technique produced monotone outputs and lowered voice tone and expressiveness. Even though FastSpeech synthesized speech faster, it still lacked naturalness and quality \cite{ren2020fastspeech}.

FastSpeech 2 addressed these issues with new features. A stronger duration predictor \cite{ren2020fastspeech} was included to improve speech temporal subtleties \cite{ren2020fastspeech}. Adding pitch and energy predictions \cite{ren2020fastspeech} improved speech prosody control and made it more natural and dynamic \cite{ren2020fastspeech}. The model's feed-forward transformers' layer normalization improved training stability \cite{ren2020fastspeech}. Most importantly, the VarPro (Variational Prosody) Loss allowed FastSpeech 2 to directly learn latent prosody representations from data, enhancing voice synthesis variability and control \cite{ren2020fastspeech}.

These developments brought new obstacles. Due to its more complex architecture, FastSpeech 2 may require more training than its predecessor. The model, like FastSpeech, relied on pre-trained autoregressive TTS model alignments, which could increase training overheads. Furthermore, its improved prosody control has drawbacks. Due to extensive modification, synthesized speech may sound unnatural or introduce artifacts.

\subsubsection{MelGAN}
MelGAN \cite{kumar2019melgan} emerged as a prominent neural vocoder capable of converting Mel-spectrograms into raw audio waveforms \cite{kumar2019melgan}. Setting itself apart from traditional vocoders like WaveNet, one of MelGAN's standout features is its significantly accelerated speech generation, proficient enough to cater to real-time speech synthesis needs. While it embraces the framework of Generative Adversarial Networks (GANs), MelGAN maintains a simplified structure, rendering it lightweight and user-friendly for both training and deployment phases. Despite its rapid generation capabilities, the quality of the audio output remains uncompromised, especially when synergized with advanced text-to-speech models like Tacotron 2 or FastSpeech \cite{kumar2019melgan}. During training, MelGAN demonstrates commendable stability compared to other GAN structures \cite{kumar2019melgan}. Notably, it sidesteps the autoregressive steps in audio generation, facilitating parallel processing and swift production of continuous audio \cite{kumar2019melgan}. Furthermore, its seamless compatibility with a variety of text-to-speech models adds to its adaptability and versatility \cite{kumar2019melgan}. Consequently, with its fusion of speed, high quality, and lightweight characteristics, MelGAN has garnered substantial attention and application within the vocoder domain.

\subsubsection{HiFi-GAN}
HiFi-GAN, with neural vocoders' noteworthy advancement, is renowned for its emphasis on high fidelity, producing audio that closely mirrors human speech in clarity and quality \cite{kong2020hifi}. This is complemented by its efficient design, tailored for real-time synthesis, making it particularly apt for applications demanding swift voice generation. Its foundation on the Generative Adversarial Network (GAN) framework \cite{kong2020hifi}, coupled with a multi-resolution approach, not only ensures stable training but also facilitates the generation of intricate waveforms \cite{kong2020hifi}. A significant edge that HiFi-GAN boasts over other vocoders is its reduced audio artifacts, ensuring the output is seamless and natural-sounding \cite{kong2020hifi}. Furthermore, its adaptability means it pairs well with a variety of text-to-speech models, enhancing its appeal in diverse speech synthesis architectures. Additionally, its robustness against overfitting, even on constrained datasets, underscores its superior design and utility in the speech synthesis domain \cite{kong2020hifi}. HiFi-GAN its sound quality even surpasses WaveNet and MelGAN in some evaluations \cite{kong2020hifi}.

\subsubsection{FastPitch}
FastPitch, which draws inspiration from the fundamental FastSpeech model, presents a paradigm-shifting methodology \cite{lancucki2021fastpitch} for the synthesis of text-to-speech (TTS). This approach deviates from conventional text-to-speech (TTS) models by prioritizing autonomy in the process of aligning audio with transcriptions \cite{lancucki2021fastpitch}. This state of self-sufficiency implies that there is no longer a dependence on external alignment models \cite{lancucki2021fastpitch}, such as Transformer TTS or Tacotron 2. It is noteworthy that the model possesses an inherent ability to forecast pitch contour \cite{lancucki2021fastpitch}, a characteristic that enhances the amplification of natural voice modulation in the resulting audio \cite{lancucki2021fastpitch}. One of the noteworthy characteristics of FastPitch is its efficacy in reducing audio artifacts, hence guaranteeing improved sound quality and accelerated training convergence \cite{lancucki2021fastpitch}. FastPitch and FastSpeech2 were trained at almost the same time \cite{lancucki2021fastpitch}.

While several models in the field rely on the distillation of mel-spectrograms with the assistance of a guiding teacher model \cite{lancucki2021fastpitch}, FastPitch diverges from this approach, simplifying the training process \cite{lancucki2021fastpitch}. An additional advantage of this system is its capability to undergo training as a multi-speaker model \cite{lancucki2021fastpitch}, hence enhancing its adaptability. FastPitch offers users a comprehensive range of options for manipulating synthesized speech, including control over prosody, speech pace, and energy levels \cite{lancucki2021fastpitch}. In order to enhance its capabilities, the model has the ability to accept either graphemes or phonemes as input, hence providing versatility in its potential applications \cite{lancucki2021fastpitch}.

One notable characteristic of the system is its capacity to perform parallel processing \cite{lancucki2021fastpitch}. The efficient generation of mel-spectrograms ensures an expedited synthesis procedure. In conclusion, FastPitch is a notable TTS system that demonstrates strength, adaptability, and effectiveness, thereby reshaping the standards within the realm of speech synthesis.

\subsubsection{Glow-TTS}

The Monotonic Alignment Search (MAS) makes Glow-TTS a cutting-edge text-to-speech model \cite{kim2020glow}. This strategy teaches the model how text and spectrogram relate, which trains a Duration Predictor for inference \cite{kim2020glow}.

The Glow-TTS encoder maps each character to a Gaussian Distribution, while the decoder uses Glow Layers to convert each spectrogram frame into a latent vector. MAS aligns these processes harmoniously, optimizing model parameters with the most likely alignment. MAS is ignored during inference, so the model only uses the duration predictor \cite{kim2020glow}.

Glow-TTS's encoder architecture is based on the TTS transformer but adds relative positional encoding and a residual Encoder Prenet link \cite{kim2020glow}. However, the decoder follows the Glow model \cite{kim2020glow}. This model trains solo and multi-speaker models well and is robust to longer sentences \cite{kim2020glow}. This durability and its capacity to infer 15 times faster than Tacotron2 make it a leader in its field \cite{kim2020glow}.

Glow-TTS's lack of auto-regressive models and external aligners is a major advantage \cite{kim2020glow}. This provides fast, high-quality performance in multi-speaker environments \cite{kim2020glow}. Its signature hard monotonic alignments guarantee focused and stable processing. Hard monotonic alignments ensure that the model stays focused on one input segment at each step, making them ideal for consistent voice synthesis \cite{kim2020glow}.

\subsubsection{OverFlow}
OverFlow is an innovative model that leverages the strengths of flow, self-regression, and Neural HMM TTS \cite{mehta2022overflow}. At its core, it integrates normalising flows with neural HMM TTS to accurately represent the intricate non-Gaussian distribution patterns observed in speech parameter trajectories \cite{mehta2022overflow}.

The primary advantage of OverFlow lies in its hybrid nature. It synergizes the strengths of classical statistical speech synthesis with contemporary neural TTS \cite{mehta2022overflow}. This synthesis results in models that need less data and fewer training updates \cite{mehta2022overflow}. Furthermore, they substantially reduce the chances of producing nonsensical output, which is a common issue attributed to neural attention failures \cite{mehta2022overflow}. Experimental data further bolsters the model's efficacy. Systems designed on the OverFlow blueprint demand fewer updates than their peers and yet consistently generate speech that rivals the naturalness of human speech \cite{mehta2022overflow}.

One of the most salient improvements introduced by OverFlow is the substitution of conventional neural attention with the neural hidden Markov model, or neural HMM. There are several reasons why this substitution is advantageous \cite{mehta2022overflow}:

Neural HMM-based models are wholly probabilistic, meaning they can be precisely trained to optimize the sequence likelihood.
The introduction of left-to-right no-skip HMMs \cite{mehta2022overflow} ensures strict adherence to the sequence, known as monotonicity \cite{mehta2022overflow}. This ensures each phonetic unit in the input is spoken in the right sequence, eliminating the challenge of non-monotonic attention \cite{mehta2022overflow}. Such issues are notorious for causing many attention-based neural TTS models to generate incoherent gibberish. Furthermore, these models usually demand vast amounts of data and numerous updates to master proper speec \cite{mehta2022overflow}h.
However, it's worth noting that merely relying on neural HMMs may not unlock their full potential due to their restrictive assumptions. Systems like those detailed in certain research publications posit that state-conditional emission distributions are either Gaussian or Laplace. These assumptions equate to an L2 or L1 loss \cite{mehta2022overflow}, respectively. Given the intricate nature of human speech, which follows a highly intricate probability distribution, these assumptions might result in subpar models \cite{mehta2022overflow}.

\subsubsection{VITS: Conditional Variational Autoencoder with Adversarial Learning for End-to-End Text-to-Speech}

The VITS model presents an advanced approach to Text-to-Speech (TTS) synthesis through its parallel architecture, incorporating several innovative techniques \cite{kim2021conditional}.

Firstly, VITS employs Normalizing Flows, a method capable of learning and modeling complex probability distributions \cite{kim2021conditional}. By applying this to their conditional prior distribution, the model can mimic the probability distribution of data more accurately, leading to audio outputs that are closer to genuine speech \cite{kim2021conditional}.

Additionally, VITS utilizes Adversarial Training \cite{kim2021conditional}. This approach involves a generator and a discriminator \cite{kim2021conditional}. The generator strives to produce audio that's as close to real data as possible, while the discriminator tries to differentiate between genuine and synthesized data \cite{kim2021conditional}. As training progresses, the generator continuously improves in an attempt to "deceive" the discriminator, thereby producing higher-quality audio \cite{kim2021conditional}. In VITS, this strategy is applied directly on the audio waveform, aiming for a more natural speech waveform output \cite{kim2021conditional}.

Beyond generating high-quality audio, VITS also emphasizes the importance of the one-to-many relationship \cite{kim2021conditional} . In natural language pronunciation, the same text can be articulated in multiple ways, influenced by factors like pitch and duration \cite{kim2021conditional}. To capture this diversity, VITS introduces a stochastic duration predictor that can synthesize speech with various rhythms from the same text input \cite{kim2021conditional}.

In essence, VITS not only focuses on producing fine-grained audio quality but also seeks to simulate the diverse pronunciations of text found in the real world. This allows it to generate rich, natural, and varied voice outputs, meeting the high standards expected of modern TTS systems.

\subsubsection{SpeechT5}

SpeechT5 has emerged as a groundbreaking framework tailored to handle a spectrum of spoken language tasks \cite{ao2021speecht5}. 

At its core, the model benefits from the strengths of a pre-trained Transformer encoder-decoder architecture \cite{ao2021speecht5}. To efficiently process both textual and auditory data, SpeechT5 seamlessly integrates pre-nets and post-nets \cite{ao2021speecht5}. Following its foundational pre-training, the model is further refined and honed to cater to specific tasks, making use of the pertinent pre/post-nets. 

Traditional models in this domain often sidelined the importance of textual data when delving into spoken language tasks \cite{ao2021speecht5}. Additionally, an excessive dependence on pre-trained speech encoders meant that the critical role of decoders was often undervalued \cite{ao2021speecht5}. Moreover, there was a conspicuous absence of a profound exploration into an encoder-decoder design that synergized both speech and text data sources \cite{ao2021speecht5}. 

Addressing these gaps, SpeechT5 draws inspiration from the esteemed T5 model and pioneers in-depth encoder-decoder pre-training founded on self-supervised learning principles \cite{ao2021speecht5}. Beyond just its architectural design, SpeechT5 introduces an innovative cross-modal vector quantization technique, creating a harmonious fusion of speech and textual data inputs \cite{ao2021speecht5}. 
Testament to its revolutionary design and capabilities, SpeechT5 not only harnesses the power of both unlabeled speech and text data but also sets new benchmarks across multiple spoken language tasks. Whether it's speech recognition or voice conversion, the model has showcased unparalleled proficiency \cite{ao2021speecht5}. Furthermore, empirical evaluations underscore the unparalleled effectiveness of the model's unique approach to joint speech-text pre-training.

\subsubsection{VALL-E}

VALL-E is a cutting-edge TTS framework that introduces an innovative approach of viewing TTS as a language modeling task. Distinct from other existing systems, what sets VALL-E apart is its use of audio encoder-decoder codes as an intermediate representation, instead of the conventional mel spectrograms. This unique design obviates the need for additional architectural engineering, pre-designed acoustic features, and fine-tuning. Emphasizing the use of a vast amount of semi-supervised data, it employs 60K hours of data for pre-training \cite{valle}, far surpassing prior pre-training endeavors. Another highlight is its powerful context-learning ability akin to GPT-3 \cite{valle}, implying that the model can comprehend and manage intricate contextual information without the need for intricate adjustments \cite{valle}.

Delving deeper into VALL-E, its diverse output emerges as a salient feature. It can generate multiple outputs for a single input text while also retaining the acoustic environment and emotions of a given acoustic cue \cite{valle}. This means that with just a 3-second recording of an unseen speaker \cite{valle}, VALL-E can synthesize high-quality voice with personalized characteristics \cite{valle}. In fact, when compared to other advanced zero-shot TTS systems \cite{valle}, VALL-E showcases significant advantages in terms of voice naturalness and speaker similarity.

In conclusion, VALL-E not only brings forth several disruptive innovations but also presents a plethora of clear advantages, paving the way for future TTS research and applications.

\subsubsection{VALL-E X}
In recent research, cross-lingual speech synthesis and foreign accents in text-to-speech synthesis have been areas of keen interest \cite{vall_ex}. Traditional cross-lingual speech synthesis models often fall short in voice quality when compared to monolingual TTS models due to data scarcity and model capacity constraints \cite{vall_ex}. Furthermore, a prevalent challenge is that cross-lingual speech synthesis frequently produces speech tinted with the accent of the source language, a problem that poses noticeable impacts across various scenarios \cite{vall_ex}.

To address these issues, researchers developed VALL-E X, a cross-lingual neural encoder-decoder language model trained on a vast amount of multilingual, multi-speaker, multi-domain non-distinct speech data \cite{vall_ex}. The central idea of this model is to utilize the source language's speech and the target language's text as cues to predict the target language's acoustic tokens \cite{vall_ex}. Distinct from traditional TTS methods, VALL-E X views TTS as a conditional language modeling task, adopting neural encoder-decoder codes as the acoustic intermediate representation \cite{vall_ex}.

The most significant advantage of VALL-E X is its capability to produce high-quality cross-lingual speech \cite{vall_ex}. It not only preserves the voice, emotion, and acoustic environment of unseen speakers but can also synthesize speech with the native accent for any speaker \cite{vall_ex}. This means it has largely addressed the issue of foreign accents in cross-lingual speech synthesis \cite{vall_ex}. Empirical tests have shown that, compared to existing baseline methods, VALL-E X stands out noticeably in terms of speaker similarity, voice quality, translation quality, voice naturalness, and human evaluations \cite{vall_ex}.

In summary, VALL-E X offers a potent tool to tackle the challenges of cross-lingual speech synthesis, significantly enhancing voice quality and naturalness. This advancement sets an important benchmark for future cross-lingual speech synthesis research.

\subsubsection{Meta-MMS}

The Massively Multilingual Speech (MMS) project is a groundbreaking initiative that aims to amplify the linguistic coverage of speech technology by 10 to 40 times, contingent on the task \cite{meta-mms}. This venture is particularly notable for the introduction of two novel datasets, MMS-lab and MMS-unlab, encompassing labeled speech data from 1,107 languages and unlabeled speech data from 3,809 languages, respectively \cite{meta-mms}. The researchers have further developed an encompassing pre-trained wav2vec 2.0 model that covers 1,406 languages \cite{meta-mms}, a singular multilingual automatic speech recognition model that caters to 1,107 languages, equivalent voice synthesis models, and a language recognition model that identifies up to 4,017 languages \cite{meta-mms}.

One of the standout features of the project is its commitment to large-scale multilingual support \cite{meta-mms}. In comparison to existing speech technologies, the models and datasets brought forward by this initiative envelop a broader range of languages, especially those that are resource-scarce and on the brink of extinction \cite{meta-mms}. A unique facet in terms of dataset collection was the utilization of publicly available religious texts recitations as the source, a distinctive avenue hitherto untapped for multilingual voice technology \cite{meta-mms}.

The advantages presented by Meta-MMS are manifold. Firstly, the sheer linguistic diversity it supports, covering over 1,000 languages \cite{meta-mms}, plays a pivotal role in the conservation and fostering of linguistic diversity. Additionally, despite the vast augmentation in linguistic coverage, the project's models continue to offer high-caliber automatic voice recognition and text-to-speech synthesis.

Addressing long-standing issues, the research mitigates the restrictive linguistic coverage that has marred voice technology. While the past decade has seen significant strides in the field, the technologies could not cater to the majority of the 7,000-plus global languages \cite{meta-mms}. The present study expands the linguistic ambit of voice technology by unveiling new datasets and models. Moreover, the extinction risk faced by numerous languages is countered by this project. The constricted linguistic span of current technologies might exacerbate extinction trends, and by bolstering support for more languages, this project serves as a bulwark for these endangered languages.

\section{Application of Transfer Learning in Various TTS Models}
\textbf{Tacotron2}: As mentioned in the literature review, Tacotron2 built upon the original Tacotron model by refining components like the encoder and attention mechanism. Research by \citeA{shen2018natural}. demonstrated Tacotron2's ability to leverage transfer learning by initializing the model with pre-trained components. This improved stability and lowered data requirements.

\textbf{FastSpeech2}: This non-autoregressive model relies on alignments from an autoregressive "teacher" model like Tacotron2. \citeA{ren2020fastspeech} showed FastSpeech2 can utilize transfer learning from the teacher to bootstrap its training. Its variance adaptor also aids adaptation to new speakers.

\textbf{VITS}: \citeA{kim2021conditional} highlighted how VITS leverages normalizing flows and adversarial training to capture complex speech distributions. Its use of normalizing flows is well-suited for adapting to new distributions via fine-tuning, enhancing few-shot transfer capabilities.

\textbf{Glow-TTS}: As a flow-based generative model, Glow-TTS can adapt its prior distribution to new datasets through fine-tuning, as noted by \citeA{kim2020glow}. Its lightweight design also enables rapid fine-tuning for transfer learning.

\textbf{OverFlow}. \citeA{mehta2022overflow} emphasized OverFlow's hybrid design integrating Neural HMM with flows. The probabilistic nature of Neural HMMs suits low-resource scenarios, providing a strong foundation for fine-tuning and transfer \cite{mehta2022overflow}.

\textbf{SpeechT5}. \citeA{ao2021speecht5} showed SpeechT5's versatility across spoken language tasks via fine-tuning its pre-trained encoder-decoder model. Its ability to perform various downstream tasks highlights SpeechT5's transfer learning aptitude.

In summary, advancements like attention mechanisms, variance adaptors, normalizing flows, Neural HMMs, and pre-training have expanded TTS models' capacity to effectively leverage transfer learning, even in low-resource settings. The literature provides encouraging evidence regarding transfer learning's potential to enable customized TTS with limited data.

\section{Conclusion}
In the domain of text-to-speech  the literature review reveals a vibrant and rapidly evolving field. The examination of key technologies, methodologies, and frameworks such as VALL-E highlights the convergence of novel approaches and cutting-edge innovations. A distinctive trend in the literature is the shift towards embracing neural architectures, large-scale data training, and cross-lingual capabilities, reflecting the broader goals of enhancing voice naturalness, speaker similarity, and linguistic diversity. Furthermore, projects such as the Meta-MMS have illuminated the potential to bridge technological gaps, particularly in underserved and endangered languages. Challenges, including the scarcity of data in cross-language voice synthesis, are being addressed through creative solutions and the development of diverse datasets. The integration of TTS as a language modeling task and the application of semi-supervised data have emerged as remarkable innovations, setting the stage for future research. This literature review underscores the dynamism and complexity of the voice cloning field, highlighting not only the achievements but also the areas ripe for exploration, experimentation, and ethical consideration. The convergence of these elements signifies a promising trajectory for voice cloning, with a clear emphasis on accessibility, quality, and the humanization of synthetic speech. It sets the stage for future work, where the synthesis of voices is not merely a technological feat but a nuanced and responsive interplay of science, art, and societal needs.

%% file: methodology.tex
\chapter{Methodology}
This section of the thesis delineates the methodological approach undertaken to conduct this research. The overarching objective is to provide a comparative analysis of transfer learning in several state-of-the-art deep learning text-to-speech models on a few-shot, low-resource, customized dataset. The methodology is designed to be thorough, systematic, and replicable, allowing for the validation and verification of the results obtained.

The choice of methodology is intrinsically tied to the research questions, which explore the potential differences in performance and efficiency among the selected models when fine-tuned with a low-resource, customized dataset. The models chosen for this study, namely Tacotron 2, VITS, Glow-TTS, FastSpeech2, OverFlow, FastPitch, and SpeechT5, were selected based on their varying architectures and unique design philosophies that hold the potential for interesting comparative outcomes.

This section is divided into several subsections that cover different aspects of the research process, including the selection of models, detailed description of each model's architecture and pipeline, and the evaluation metrics used to assess their performance. Furthermore, the processes of data collection, data preprocessing, and model fine-tuning will also be discussed.

Additionally, the crucial aspect of model parameter selection will be addressed. This section will cover the key parameters chosen for each model, the rationale behind their selection, and the importance of these parameters to the model's overall performance. The importance of selecting suitable parameters cannot be overstated as it significantly influences the performance of each model.

The methodology aims to provide comprehensive and clear explanations of the procedures followed, enabling an understanding of how the research was conducted, and thus, allowing an accurate assessment of its validity and reliability. This ensures that the conclusions drawn are robust, reliable, and contribute significantly to the field of voice cloning using deep learning models.

In the course of our experimentation, we leveraged multiple open-source text-to-speech toolkits to conduct relevant tests and evaluations. The primary toolkits and repositories we made use of are as follows:
\begin{itemize}
  \item NVIDIA's Tacotron2\footnote{https://github.com/NVIDIA/tacotron2} offered an attention mechanism-based sequence-to-sequence voice synthesis model, giving another powerful baseline for our experiments.
  \item The VITS model\footnote{https://github.com/jaywalnut310/vits}, provided by jaywalnut310, represents an efficient TTS system that integrates several cutting-edge technologies.
  \item SpeechT5 developed by Microsoft\footnote{https://github.com/microsoft/SpeechT5}, blends the T5 architecture into voice tasks, providing a valuable benchmark for our comparative experiments.
  \item We also employed FastSpeech2 by ming024\footnote{https://github.com/ming024/FastSpeech2}, a non-autoregressive voice synthesis method based on the Transformer.
    \item The TTS toolkit by coqui-ai\footnote{https://github.com/coqui-ai/TTS}, which encompasses a variety of TTS models and utilities, facilitated our experimental processes.
    \item An additional version of VITS by Plachtaa\footnote{https://github.com/Plachtaa/VITS-fast-fine-tuning} was also utilized in certain experimental settings.
\end{itemize}

By employing the aforementioned tools and repositories, we were able to extensively assess the performance of different models on our specific dataset and pinpoint the optimal solution.

%  Model Selection, Architectures and Pipelines

\section{Model Selection, Architectures and Pipelines}

In order to provide a comprehensive comparative analysis of transfer learning in text-to-speech synthesis, a wide array of models were selected, each embodying unique architectural features, training paradigms, and strengths. The following sub-sections provide an overview of these models and the reasoning behind their selection for this study.

%%% ------------------ Tacotron2--------------------------%%%%%
%%% ------------------ Tacotron2--------------------------%%%%%
%%% ------------------ Tacotron2--------------------------%%%%%
%%% ------------------ Tacotron2--------------------------%%%%%
% Tacotron2
\subsection{Tacotron2}
Tacotron2 (Shen et al., 2018) is a highly-regarded model in the field of TTS. Leveraging an end-to-end system, Tacotron2 eliminates the need for intricate text and audio processing pipelines. Its utilization of a recurrent sequence-to-sequence architecture with attention allows us to explore the impact of such features on the model's ability to adapt to new, low-resource scenarios via transfer learning.

% Tacotron2 Architecture
\subsubsection{Tacotron2 Architecture}
Tacotron 2, as an end-to-end generative text-to-speech model, comprises of two main parts: a recurrent sequence-to-sequence feature prediction network with attention (referred to as Tacotron) and a modified WaveNet model acting as a vocoder.

% Tacotron文章架构图
\begin{figure}[ht]
    \centering
    \includegraphics[width=0.5\textwidth]{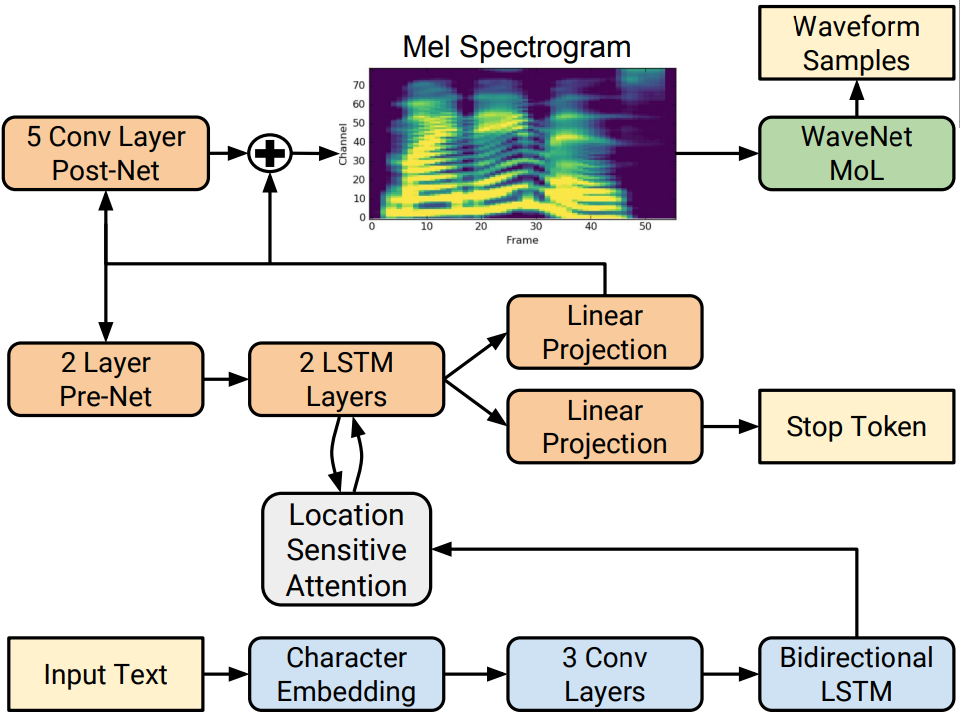}
    \caption[Architecture of the Tacotron2 Model]{Tacotron2 model architecture. Figure extracted from \protect\cite{shen2018natural}.}
    \label{fig:tacotron2_arch}
\end{figure}

\textbf{Character Encoding}: The initial step in Tacotron 2 involves the conversion of input text into a sequence of linguistic features. These features, often phonemes or graphemes, are then passed into an embedding layer, which translates these discrete symbols into continuous vectors.
    
\textbf{Encoder}: The continuous vectors are then fed into the encoder, which in the case of Tacotron 2, is a stack of three convolutional layers followed by a single bidirectional LSTM layer. This encoder processes the input sequence into a higher-level representation that encapsulates the context of each character in the text.
    
\textbf{Attention and Decoder}: Tacotron 2 uses a location-sensitive attention mechanism to align the encoder outputs with the decoder inputs. The decoder, a stack of LSTM layers, takes this alignment information along with the previously predicted spectrogram frame and generates the next spectrogram frame.

\textbf{Post-net}: After the decoder generates a mel-spectrogram, it is passed through a post-net which is a stack of convolutional layers. This post-net refines the mel-spectrogram prediction, enhancing the output's quality.
    
\textbf{WaveNet Vocoder}: Finally, the predicted mel-spectrogram is converted into time-domain waveforms using the WaveNet vocoder. The WaveNet vocoder uses the predicted mel-spectrogram as a local condition to generate high-fidelity speech.

% \subsubsection{Implementation Details}

% In the scope of this project, Tacotron2 was implemented using the PyTorch library, a popular open-source machine learning framework that provides significant flexibility and efficiency. The model was trained from scratch using the LJSpeech dataset, which is a widely used and reliable dataset for training voice cloning models.

% \subsubsection{Specific Considerations}

% Despite its high-quality outputs, it is important to note that Tacotron2 is computationally intensive. The training process requires significant computational resources, including high-performing GPUs and a large amount of memory. This requirement presents a challenge in terms of computational efficiency, particularly for large-scale applications.

%%% ------------------ Tacotron2 END--------------------------%%%%%
%%% ------------------ Tacotron2 END--------------------------%%%%%
%%% ------------------ Tacotron2 END--------------------------%%%%%

%%% ------------------ FASTSPEECH2 START--------------------------%%%%%
%%% ------------------ FASTSPEECH2 START--------------------------%%%%%
%%% ------------------ FASTSPEECH2 START--------------------------%%%%%

\subsection{FastSpeech2}

In the continuously evolving realm of text-to-speech (TTS) synthesis, the challenge of the one-to-many mapping problem has been a persistent hurdle. FastSpeech2 has paved a transformative path to address this issue. Instead of relying on the over-simplified output typically derived from teacher models, it prioritizes the use of ground-truth targets during training, ensuring a synthesis process that is both genuine and authentic. Additionally, FastSpeech2 uniquely integrates a broader range of speech variation data as conditional inputs. These variations, which envelop crucial speech elements such as pitch, energy, and a refined precision in duration, are extracted directly from the speech waveform. This approach allows FastSpeech2 to adeptly use them during the training phase and employ the predicted values during inference. The results of these innovative changes are nothing short of remarkable. FastSpeech2 has showcased a training speed that surpasses its predecessor, FastSpeech, by a factor of three \cite{ren2020fastspeech}. 

\subsubsection{FastSpeech2 Architecture and Pipeline }

\begin{figure}[ht]
    \centering
    \includegraphics[width=1.0\textwidth]{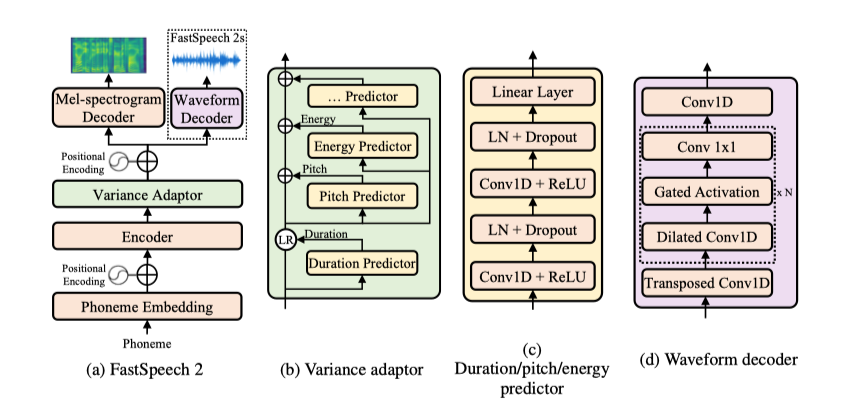}
    \caption[Architecture of the FastSpeech2 Model]{The diagram showcases the flow from phoneme input through phoneme embedding, positional encoding, and the encoder to the variance adaptor, and finally to the mel-spectrogram decoder. The model integrates specific variance information such as duration, pitch, and energy at various stages to produce a high-quality mel-spectrogram output, which is then processed by a vocoder to produce the final audio waveform Figure extracted from \protect\cite{ren2020fastspeech}.}
    \label{fig:fastSpeech2_arch}
\end{figure}

\textbf{Phoneme Embedding}: The model begins its processing by taking phonemes as input. These phonemes undergo a transformation via a Phoneme Embedding layer, converting them into continuous-valued vectors. This conversion facilitates better representation and processing in subsequent layers \cite{ren2020fastspeech}.

\textbf{Positional Encoding Addition}: Once we have our phoneme embeddings, it's essential to provide the model with positional information. Positional Encoding is integrated into these embeddings, ensuring each position (like characters or words in the text) receives a unique encoding, giving the model a clear sense of sequence order \cite{ren2020fastspeech}.

\textbf{Encoder}: These positionally encoded phoneme embeddings are then channeled through the encoder. This stage refines these embeddings, yielding hidden representations that encapsulate both the phonemic and positional nuances of the input \cite{ren2020fastspeech}.

\textbf{Variance Adaptor}: The model then employs a Variance Adaptor, which infuses the hidden sequence with additional variance details, namely duration, pitch, and energy. These metrics are pivotal as they contribute to the naturalness and expressivity of the speech output \cite{ren2020fastspeech}.

\textbf{Second Positional Encoding}: Given the richness and complexity of the hidden representations after the Variance Adaptor, there's a need to reintroduce positional encoding. This step is crucial as, through multiple transformations, the initial positional data may have become diluted. Re-applying positional encoding ensures a sharp, unambiguous retention of sequence order as the model proceeds \cite{ren2020fastspeech}.

\textbf{Mel-spectrogram Decoder}: Armed with these comprehensive representations, the model progresses to the mel-spectrogram decoder. Here, it translates the processed sequence into a mel-spectrogram, a visual representation of the spectrum of frequencies as they vary with time \cite{ren2020fastspeech}. 

\textbf{Waveform Decoder}: The last step before obtaining the audio output involves the waveform decoder. Its primary function is to transform the mel-spectrogram into raw audio waveforms, readying them for playback or storage as audio files \cite{ren2020fastspeech}.

By fusing these carefully designed components and steps, FastSpeech2 establishes itself as a formidable solution in the TTS domain, promising high-quality, natural-sounding speech outputs \cite{ren2020fastspeech}.

%%% ------------------ FASTSPEECH2 END--------------------------%%%%%
%%% ------------------ FASTSPEECH2 END--------------------------%%%%%
%%% ------------------ FASTSPEECH2 END--------------------------%%%%%

%%% ------------------ FastPitch START--------------------------%%%%%
%%% ------------------ FastPitch START--------------------------%%%%%
%%% ------------------ FastPitch START--------------------------%%%%%
%%% ------------------ FastPitch START--------------------------%%%%%
\subsection{FastPitch}
FastPitch's incorporation of two feed-forward Transformer (FFTr) stacks underscores its innovative architecture \cite{lancucki2021fastpitch}. This design ensures an enhanced processing speed compared to traditional recurrent models. The dual FFTr stacks allow for efficient parallel processing of input data, significantly reducing the time required to generate outputs. Such an architecture not only improves computational efficiency but also aids in capturing intricate patterns in the data. By leveraging the power of FFTr stacks, FastPitch ensures robustness and precision, making it an optimal choice for real-time voice synthesis tasks \cite{lancucki2021fastpitch}. Furthermore, this dual-stack mechanism can effectively handle longer sequences, ensuring consistency and clarity in the synthesized voice outputs \cite{lancucki2021fastpitch}. FastPitch and FastSpeech2 were developed at the same time \cite{lancucki2021fastpitch}.

\subsubsection{FastPitch Architecture and Pipeline}

\begin{figure}[ht]
    \centering
    \includegraphics[width=1.0\textwidth]{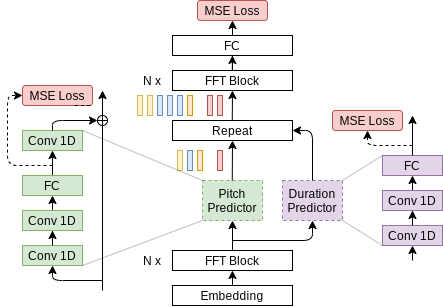}
    \caption[Architecture of the FastPitch Model]{Figure: FastPitch Architecture Overview. This diagram illustrates the integration of two feed-forward Transformer (FFTr) stacks, along with the pitch predictor and duration predictor modules. The entire structure emphasizes parallel processing, enabling efficient and simultaneous transformation of input text to mel-spectrograms \protect\cite{lancucki2021fastpitch}.}
    \label{fig:fastpitch}
\end{figure}

FastPitch employs a sophisticated feedforward Transformer methodology \cite{lancucki2021fastpitch}, meticulously designed to transmute raw textual inputs into mel-spectrograms. Remarkably, its design emphasizes parallelism, ensuring that each character within the input text is processed concurrently. This efficient design paradigm ensures that a comprehensive mel-spectrogram is crafted seamlessly within a single forward pass \cite{lancucki2021fastpitch}.

Delving deeper into its architecture, FastPitch is anchored upon a bidirectional Transformer structure, commonly recognized as a Transformer encoder \cite{lancucki2021fastpitch}. This foundational component is adeptly complemented with both pitch and duration predictors, emphasizing the model's adaptability and specificity. As input sequences make their journey through the preliminary N Transformer segments for encoding, they are not merely processed but are enriched with nuanced pitch data, subsequently experiencing discrete upsampling \cite{lancucki2021fastpitch}. This enriched data is then channeled through another set of N Transformer units \cite{lancucki2021fastpitch}. Here, the process emphasizes refinement, smoothing out any irregularities in the upsampled signal and culminating in the generation of a polished mel-spectrogram.

FastPitch is inspired by FastSpeech and predominantly incorporates two feed-forward Transformer (FFTr) stacks \cite{lancucki2021fastpitch}. The first FFTr stack processes the input tokens, while the second manages the output frames. Given an input sequence of lexical units denoted by x and a target sequence of mel-scale spectrogram frames represented by y, the first FFTr stack \cite{lancucki2021fastpitch} generates a hidden representation, symbolized as h. This representation, h, then facilitates the prediction of both duration and average pitch for each character via a 1-D CNN \cite{lancucki2021fastpitch}.

The pitch is adjusted to align with the dimensionality of the hidden representation, h, and then added to it. This modified sum, denoted as g, undergoes discrete upsampling before being input into the output FFTr \cite{lancucki2021fastpitch}, yielding the final mel-spectrogram sequence. The equation for this can be represented as:
g = h + \text{PitchEmbedding}(p), $\hat{y} = \text{FFTr}(g)$.

During training, the model employs the actual values of p and d, while predictions are used during inference. The goal is to minimize the mean-squared error (MSE) between predicted and true values \cite{lancucki2021fastpitch}.

To encapsulate, FastPitch's architecture is a testament to the harmonious integration of Transformer prowess, alongside dedicated pitch and duration prediction modules. This symbiosis ensures the creation of a formidable, highly efficient mechanism for text-to-spectrogram conversion.

%%% ------------------ FastPitch END--------------------------%%%%%
%%% ------------------ FastPitch END--------------------------%%%%%
%%% ------------------ FastPitch END--------------------------%%%%%
%%% ------------------ FastPitch END--------------------------%%%%%

%%% ------------------ GLOW-TTS--------------------------%%%%%
%%% ------------------ GLOW-TTS--------------------------%%%%%
%%% ------------------ GLOW-TTS--------------------------%%%%%
%%% ------------------ GLOW-TTS--------------------------%%%%%

\subsection{Glow-TTS}

The Glow-TTS model was selected due to its unique structure that includes a text encoder, duration predictor, and flow-based decoder. It has shown strong performance in terms of the quality of synthesized speech and the ability to maintain natural prosody and rhythm of speech. The central algorithm implemented in the Glow-TTS model is the Monotonic Alignment Search. This algorithm significantly influences the model's ability to generate synthesized speech, thus making it a critical component for evaluation. To effectively assess the impact and performance of the Monotonic Alignment Search, we have opted to select the Glow-TTS model for this study. The model's structure also makes it less complex than other models, making it a good choice for an initial comparative analysis in this study \cite{kim2020glow}.

\subsubsection{Glow-TTS Architecture and Pipeline}
\begin{figure}[ht]
    \centering
    \includegraphics[width=1.0\textwidth]{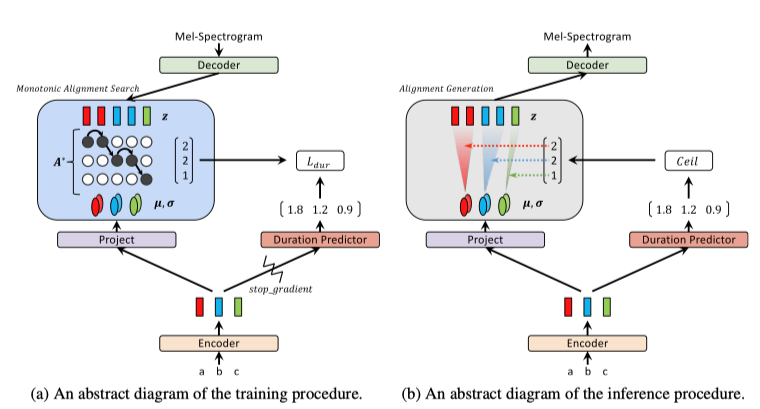}
     \caption[Architecture of the Glow-TTS Model]{The figure depicts the major components and information flow within the Glow-TTS model, including the text encoding, duration prediction, Monotonic Alignment Search process, and speech generation. It provides a visual representation of how Glow-TTS processes text input to produce synthesized speech. Figure extracted from \protect\cite{kim2020glow}.}
    \label{fig:glowtts_arch}
\end{figure}

The Glow-TTS model is composed of three main parts: a text encoder, duration predictor, and a flow-based decoder.

\textbf{Text Encoder}: This transforms the input text sequence into a continuous hidden representation. This hidden representation contains information about the phonemes input and its context within the input sequence \cite{kim2020glow}.

\textbf{Duration Predictor}: This predicts the duration of each input phoneme. The predicted duration aligns the time each input phoneme expands with the target speech time \cite{kim2020glow}.

\textbf{Flow-Based Decoder}: This transforms the outputs of the text encoder and duration predictor into the target speech. The decoder achieves this by sampling a latent variable from the prior distribution and transforming this latent variable into a mel-spectrogram \cite{kim2020glow}.

The process of the Glow-TTS model can be split into two phases: the training phase and the inference phase.

\textbf{Training Phase}:
The input text is processed through the text encoder to get a hidden representation of each phoneme. These representations are input into the duration predictor to predict the duration of each phoneme. The Monotonic Alignment Search (MAS) algorithm is used to find the most probable hard monotonic alignment \cite{kim2020glow}. The alignment and hidden representation are input into the prior encoder to get the statistical properties (mean and variance) of the prior distribution. A latent variable is sampled from the prior distribution and transformed into a mel-spectrogram by the flow decoder \cite{kim2020glow}.

\textbf{Inference Phase}:
The process during the inference phase is identical to the training phase, except that the durations used to get the alignment are those predicted by the duration predictor, and not the actual durations \cite{kim2020glow}.

%%% ------------------ GLOW-TTS END--------------------------%%%%%
%%% ------------------ GLOW-TTS END--------------------------%%%%%
%%% ------------------ GLOW-TTS--------------------------%%%%%
%%% ------------------ GLOW-TTS END --------------------------%%%%%

%%% ------------------ VITS START--------------------------%%%%%
%%% ------------------ VITS--------------------------%%%%%
%%% ------------------ VITS--------------------------%%%%%
%%% ------------------ VITS--------------------------%%%%%
% VITS part
\subsection{VITS}
The VITS (Variational Inference with adversarial learning for end-to-end Text-to-Speech) model  \cite{kim2021conditional} is selected in this study due to its unique structure and methodology which includes the application of variational inference and normalizing flows. VITS or Variational Inference Text-to-Speech operates by capturing the multimodal nature of speech data via variational inference.

VITS stands out among TTS models due to its ability to generate high-quality speech samples \cite{kim2021conditional}. Its structure integrates generative modeling with adversarial training, offering robust performance, and versatility. Moreover, it is equipped to handle complex data distributions, something inherently present in speech data. Given these capabilities, it has been selected as part of this study’s comparative analysis.

\subsubsection{VITS Architecture and Pipeline}

% VITS文章架构图
\begin{figure}[ht]
    \centering
    \includegraphics[width=1.0\textwidth]{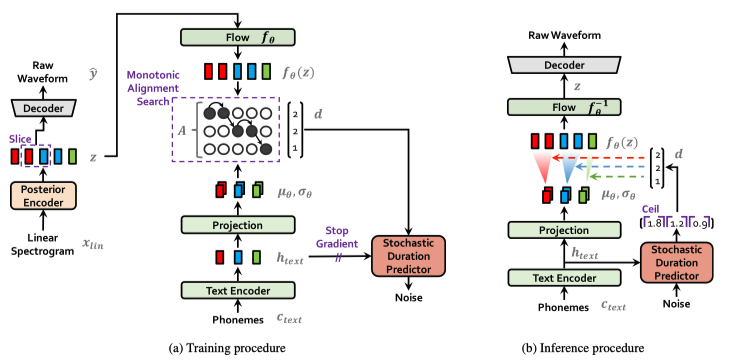}
    \caption[Architecture of the Variational Inference Text-to-Speech (VITS) Model]{The figure visualizes the detailed components and flow of information within the model, including the text encoding, prior encoding, sampling process, and speech generation. The structure showcases how the VITS model processes text input to generate synthesized speech. Figure extracted from \protect\cite{kim2021conditional}.}
    \label{fig:vits_arch}
\end{figure}

The training procedure and architecture of the VITS model can be described as a series of interconnected components, each performing a specific function in the pipeline. It can be outlined as follows:

\textbf{Data Preparation}: VITS model starts with the extraction of linear and mel spectrograms from speech samples. The linear spectrogram serves as the input to the posterior encoder to be transformed into latent variables, while the mel spectrogram is used to compute reconstruction loss \cite{kim2021conditional}. 

\textbf{Text Encoding}: Phonemes extracted from the input text are fed into a text encoder and transformed into a hidden representation, capturing helpful characteristics of the input text for subsequent speech synthesis \cite{kim2021conditional}. 

\textbf{Linear Projection Layer}: The output of the text encoder is processed by a linear projection layer that converts the hidden text representation into means and variances used for constructing the prior distribution \cite{kim2021conditional}.

\textbf{Posterior Encoder}: The linear spectrogram, after processing through the posterior encoder, yields latent variables that are sliced and fed into the normalizing flow. The posterior encoder is responsible for converting a sequence of Gaussian noise into two random variables, which are used to represent an approximate posterior distribution.  \cite{kim2021conditional}.

\textbf{Flow}: Referring to the Normalizing Flow, it is a powerful generative model capable of learning complex data distributions. In this model, the flow is applied to the prior encoder to enhance its representational power for the prior distribution, generating more realistic samples \cite{kim2021conditional}.

\textbf{Alignment Estimation}: Monotonic Alignment Search (MAS) is used to determine the alignment that maximizes the log-likelihood of the latent variables, equivalent to finding the alignment that maximizes the Evidence Lower BOund (ELBO) \cite{kim2021conditional}.

\textbf{Stochastic Duration Predictor}: This component estimates the duration distribution of the phonemes \cite{kim2021conditional}.

\textbf{Adversarial Training}: To improve synthetic quality, the model undergoes adversarial training, employing a discriminator to distinguish between real and generated speech samples. Concurrently, the generator (the model) tries to generate speech samples that the discriminator can't differentiate \cite{kim2021conditional}.

\textbf{Optimization}: The model is trained to minimize reconstruction loss and adversarial loss, and to maximize the log-likelihood of the latent variables \cite{kim2021conditional}.

%%% ------------------ VITS END--------------------------%%%%%
%%% ------------------ VITS END--------------------------%%%%%
%%% ------------------ VITS END--------------------------%%%%%
%%% ------------------ VITS END--------------------------%%%%%

%%% ------------------ SpeechT5 START--------------------------%%%%%
%%% ------------------ SpeechT5 START--------------------------%%%%%
%%% ------------------ SpeechT5 START--------------------------%%%%%
%%% ------------------ SpeechT5 START--------------------------%%%%%

\subsection{SpeechT5}
Choosing SpeechT5 as a model for spoken language processing tasks stems from its multifaceted capabilities. It stands as a unified-modal framework with immense power to handle a plethora of tasks such as automatic speech recognition, speech synthesis, speech translation, voice conversion, speech enhancement, and speaker identification \cite{ao2021speecht5}. Its versatility is unparalleled, especially when considering that many models are often designed specifically for a single task, while SpeechT5 has the dexterity to span across multiple.

A distinct feature of SpeechT5 is its ability to train without labels \cite{ao2021speecht5}. It's grounded on the Transformer encoder-decoder paradigm, allowing it to be pre-trained without the necessity for labeled data. This not only reduces the demands of training data but enhances the model's adaptability in multitask scenarios.

\subsubsection{SpeechT5 Architecture and Pipeline}

% VITS文章架构图
\begin{figure}[ht]
    \centering
    \includegraphics[width=1.0\textwidth]{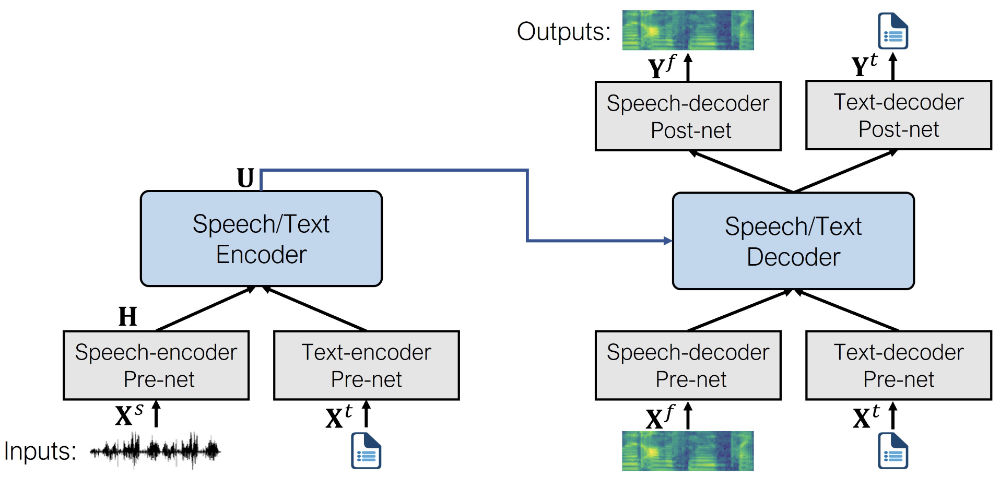}
    \caption[Architecture of the SpeechT5 Model]{Architecture of SpeechT5: A Unified-Modal Framework for Spoken Language Processing. The diagram showcases the integration of the Transformer encoder-decoder core with specialized pre-nets and post-nets, facilitating the model's adaptability across various tasks such as speech recognition, synthesis, translation, and more. Figure extracted from \protect\cite{ao2021speecht5}.}
    \label{fig:speecht5}
\end{figure}

The architecture of SpeechT5 pivots around a pre-trained Transformer encoder-decoder core, ensuring proficiency in managing both speech and text data. The inclusion of "pre-nets" serves the purpose of transforming the input, be it text or speech, into hidden representations that the Transformer can utilize. On the other side, "post-nets" take the Transformer's output and reformat it back into text or speech, ensuring a coherent output format \cite{ao2021speecht5}.

A noteworthy aspect is the model's adaptability. Post its pre-training phase, the entire encoder-decoder structure can be fine-tuned to cater to individual tasks \cite{ao2021speecht5}. This fine-tuning leverages specific pre-nets and post-nets tailored for the task in question, which means SpeechT5 can offer specialized solutions while retaining a consistent core architecture.

In essence, SpeechT5 offers not just a robust solution for spoken language processing but presents an adaptable and comprehensive approach, making it a prime choice for a range of applications.

%%% ------------------ SpeechT5 END--------------------------%%%%%
%%% ------------------ SpeechT5 END--------------------------%%%%%
%%% ------------------ SpeechT5 END--------------------------%%%%%
%%% ------------------ SpeechT5 END--------------------------%%%%%

%%% ------------------ OverFLow START--------------------------%%%%%
%%% ------------------ OverFLow START--------------------------%%%%%
%%% ------------------ OverFLow START--------------------------%%%%%
%%% ------------------ OverFLow START--------------------------%%%%%
\subsection{OverFLow}
In the quest to delve deeper into text-to-speech systems, the choice of using OverFlow stands out for several pivotal reasons:

OverFlow integrates the Neural HMM, positioning itself as a distinctive and promising model within the TTS domain. Given the contemporary technological landscape, the Neural HMM offers intriguing possibilities for transfer learning in situations characterized by limited samples and resources \cite{mehta2022overflow}. Our ambition is to further probe and validate the efficacy of Neural HMM in such specific scenarios .

Beyond merely incorporating Neural HMM, OverFlow amalgamates elements of flow techniques with autoregression, bestowing a more nuanced and comprehensive depiction for TTS \cite{mehta2022overflow}. This composite approach stands to bolster the accuracy and stability of the model.

To elaborate further, OverFlow layers normalising flows atop the foundational neural HMM TTS. This design choice equips the model to aptly describe the non-Gaussian distribution intrinsic to speech parameter trajectories \cite{mehta2022overflow}. This suggests that the model is adept at precisely capturing the intricate patterns and nuances present in speech data.

In summary, the primary impetus behind our selection of OverFlow is its fusion of multiple cutting-edge techniques, particularly the Neural HMM, coupled with our intrigue about its potential in low-resource transfer learning scenarios \cite{mehta2022overflow}.

\begin{figure}[ht]
    \centering
    \includegraphics[width=1.0\textwidth]{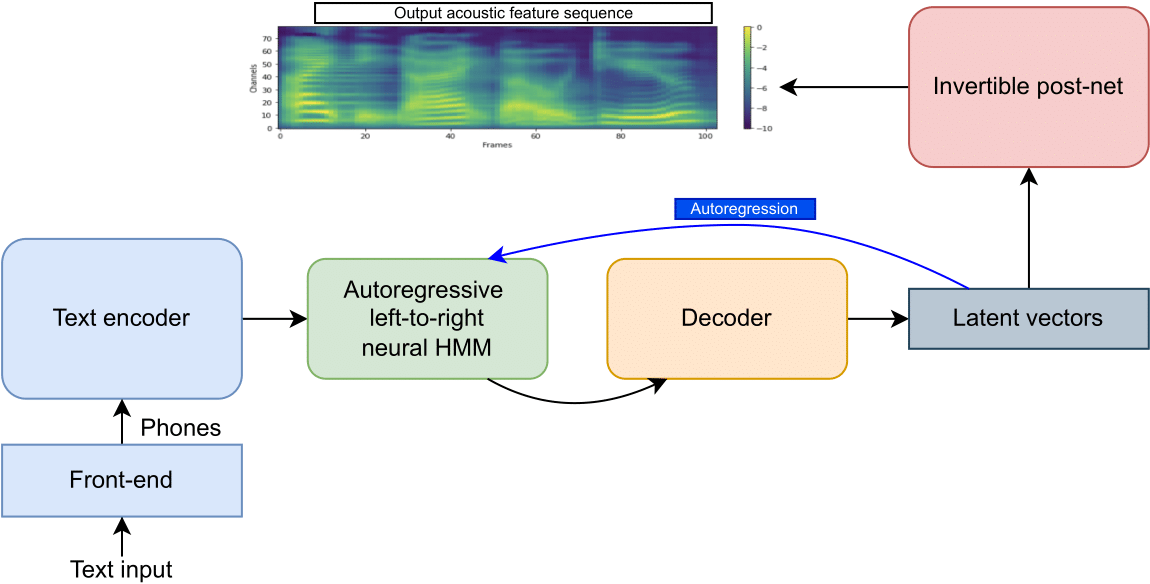}
    \caption[Architecture of the OverFlow Model]{Architecture of the OverFlow Text-to-Speech System - An integrated representation showcasing the synergy of Neural HMM with normalizing flows and autoregression, designed to accurately capture non-Gaussian speech parameter trajectories \protect\cite{mehta2022overflow}.}
    \label{fig:overflow}
\end{figure}

\subsubsection{OverFlow Architecture and Pipeline}

\textbf{Text Encoder}: In the context of TTS acoustic modeling, the neural HMM encoder transforms input vectors, such as phone embeddings, into a sequence of vectors \(h_{1:N}\) that represent \(N\) states in a left-to-right, no-skip Hidden Markov Model (HMM). Notably, each input symbol maps to a fixed number of vectors, sometimes two states per symbol \cite{mehta2022overflow}.

\textbf{Autoregressive Left-to-Right Neural HMM Workflow}: 
Neural HMM characterizes sequences, with each frame typically assuming a Gaussian distribution. The technique of Normalizing Flows takes this base distribution and applies an invertible nonlinear transformation \(f\), evolving it into a more intricate target distribution. This transformation can be mathematically represented as \(X = f(Z; W )\), where \(X\) denotes the target distribution, \(Z\) signifies the source distribution, and \(W\) encapsulates the parameters guiding the transformation. The term "left-to-right" implies a system progression from the sequence's start to its end without omitting any intermediate steps. In many scenarios, each input symbol might translate into multiple state vectors \cite{mehta2022overflow}.

The primary roles of Neural HMMs are probabilistic modeling and serving as an alternative to traditional neural attention. Neural HMMs provide a distributional description of discrete-time sequences, often assuming a Gaussian distribution for each frame. They offer a framework that aligns input values with output observations, standing as an alternative to traditional neural attention \cite{mehta2022overflow}.

\textbf{Decoder}: 
The decoder within the neural HMM (Hidden Markov Model) produces outputs from the provided inputs. Its primary objective is to generate the parameters of the emission distribution and the transition probability. This dual capability aids in delineating how the HMM transitions to the impending state and the output frame that should be produced for each timestep \cite{mehta2022overflow}.

The neural HMM decoder accepts two primary inputs: a state-defining vector \(h_{st}\), corresponding to the HMM state \(st\) at timestep \(t\), and the previously generated acoustic frames \(x_{1:t-1}\), making the model inherently autoregressive. The decoder's dual outputs are the parameters \(\theta_t\) of the emission distribution and the transition probabilities \(\tau_i\). The parameter \(\theta_t\) describes the next output frame \(xt\)'s probability distribution. Most neural HMMs postulate that \(xt\) adheres to a multivariate Gaussian distribution with a diagonal covariance matrix. Under this assumption, the parameter \(\theta_t\) bifurcates into two vectors that respectively correspond to \(xt\)'s element-wise mean and standard deviation \cite{mehta2022overflow}.

The transition probability \(\tau_t\) outlines the probability of the HMM transitioning to a subsequent state. If this transition manifests, then \(st+1 = st + 1\); otherwise, \(st+1 = st\). Synthesis traditionally commences from \(s1 = 1\) and concludes when \(st = N + 1\). To meet the Markov assumption on the concealed states \(s_{1:T}\), the decoder's outputs must remain independent of the input vectors from previous time steps \(h_{1:t-1}\). This independence enables training the model to maximize training data's log-likelihood using classical HMM's standard algorithms \cite{mehta2022overflow}.

\textbf{Invertible Post-net}: 
The invertible post-net primarily enhances output quality and ensures compatibility with maximum likelihood training. Many autoregressive TTS systems use a greedy, sequential output generation method during synthesis. To improve this output, a post-net with non-causal CNNs refines the sequence generated by the autoregressive model.

Traditional post-net architectures aren't compatible with neural HMMs' maximum likelihood training. But by passing the model output through an invertible post-net, the entire model can be trained to maximize the exact sequence likelihood.

Normalizing flows transform simple latent distributions into more complex target distributions via an invertible nonlinear transformation. The ability to reverse this process allows for the computation of the log-probability of any observed result using the change-of-variables formula. Even when the transformation is slightly non-linear, complex distributions can emerge by chaining several transformations together, often labeled as an invertible neural network \cite{mehta2022overflow}.

%%% ------------------ OverFLow END--------------------------%%%%%
%%% ------------------ OverFLow END--------------------------%%%%%
%%% ------------------ OverFLow END--------------------------%%%%%
%%% ------------------ OverFLow END--------------------------%%%%%

%%% ------------------ Model 讲解 END--------------------------%%%%%
%%% ------------------ Model 讲解 END--------------------------%%%%%

%%% ------------------ Data Collection and Analysis START --------------------------%%%%%
%%% ------------------ Data Collection and Analysis START--------------------------%%%%%

% Data Collection and Analysis part
\section{Data Collection}
For the purpose of examining the model's ability to generalize on random small datasets \cite{rothauser1969ieee}, I read out and record 7 lists from The Harvard Sentences. The Harvard Sentences, sometimes referred to as the Harvard Lines, comprise a set of 720 standardized phrases. These phrases are organized into 10 distinct lists and are specifically designed for testing on systems like Voice over IP, mobile phones, and other telecommunication devices. Phonetically balanced, these sentences reflect the frequency of phoneme occurrences in the English language. As part of the IEEE Recommended Practice for Speech Quality Measurement, these sentences are categorized under the 1965 Revised Speech Balanced Sentence List. Their consistent and standard nature makes them an invaluable tool in the fields of telecommunications, speech, and acoustics research \cite{rothauser1969ieee}.

The dataset contains individual sentences like "The birch canoe slid on the smooth planks." To enrich the training set with extended sentences, I merged pairs of these statements. To test transfer learning from a few-shot dataset, my dataset ultimately encompasses 49 sentences, totaling 341 unique words.

I have also created a "metadata.txt" file, which contains the dataset's transcript to provide labels for training and testing.

\section{Data Analysis}
The total audio duration of all sentences is approximately 266 seconds, with an average duration of 5 seconds per sentence, ranging from 3 to 9 seconds. I employed a sampling rate of 22050Hz, mono audio, with 16 bits per sample, which is a common choice in voice cloning research due to the balance it provides between audio quality and computational efficiency.

\subsection{Dataset Distribution Analysis}
To guarantee the randomness of my dataset and subsequently test the impact of a random dataset on the model's transfer learning capabilities, I compared both the "text length vs. mean audio duration" and "text length vs. number of instances" in my dataset to those in the LJSpeech \cite{ljspeech17} dataset. 

LJSpeech is a renowned public domain dataset that stands out in the realm of voice and speech research. Comprising concise audio recordings, the heart of this dataset lies in its meticulously curated content: passages read aloud by a singular female speaker, sourced from a diverse collection of seven non-fiction English books. These readings encompass a vast array of topics, spanning from profound philosophical discourses and poetic expressions to rigorous scientific analyses \cite{ljspeech17}.

Beyond the sheer breadth of content, the dataset's magnitude is equally impressive, amassing to an aggregate of approximately 24 hours of speech from native speakers \cite{ljspeech17}. The dataset contains over 13,000 short audio clips and corresponding transcripts, with the audio clips totaling more than 24 hours in length \cite{ljspeech17}. One of LJSpeech's defining features is its pairing mechanism; each audio snippet is meticulously matched with its corresponding textual transcript \cite{ljspeech17}. This characteristic not only anchors the dataset in the realm of supervised machine learning but also makes it an invaluable asset for the intricate processes of speech synthesis, especially for the training of cutting-edge Text-To-Speech (TTS) models.

\begin{figure}[ht]
    \centering
    \includegraphics[width=0.5\textwidth]{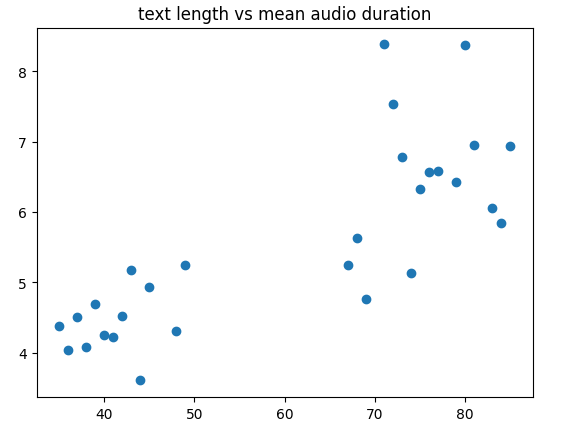}
    \caption[My Dataset Text Length vs Mean Audio Duration]{My dataset text length vs mean audio duration}
    \label{fig:my_dataset_text_length_vs_audio_duration}
\end{figure}

\begin{figure}[ht]
    \centering
    \includegraphics[width=0.5\textwidth]{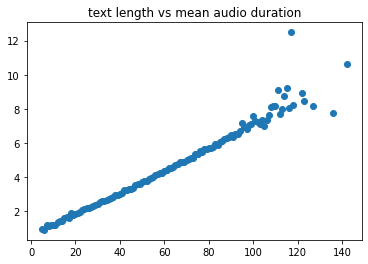}
    \caption[LJSpeech Dataset Text Length vs Mean Audio Duration]{LJSpeech dataset text length vs mean audio duration}
    \label{fig:ljspeech_dataset_text_length_vs_audio_duration}
\end{figure}

In speech synthesis research, the relationship between text length and the number of instances within a dataset provides crucial insights into its composition and potential biases. This "Text Length vs Number of Instances" analysis elucidates the distribution of various sentence lengths present in the dataset. A balanced distribution ensures that the model is exposed to a diverse range of sentence structures and lengths during training, facilitating better generalization during real-world applications. In the context of our study, we delve into this relationship not only to understand our dataset's inherent characteristics but also to ensure its robustness in training models, particularly when evaluating the effects of transfer learning on randomly distributed data.

\begin{figure}[ht]
    \centering
    \includegraphics[width=0.5\textwidth]{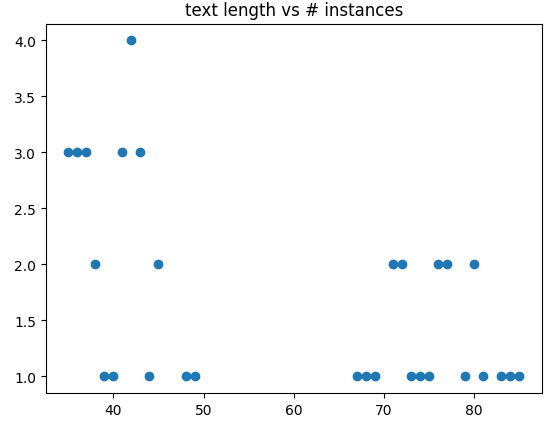}
    \caption[My Dataset Text Length vs Number of Instances]{My Dataset Text Length vs Number of Instances}
    \label{fig:my_text_length_vs_number_of_Instances}
\end{figure}

\begin{figure}[ht]
    \centering
    \includegraphics[width=0.5\textwidth]{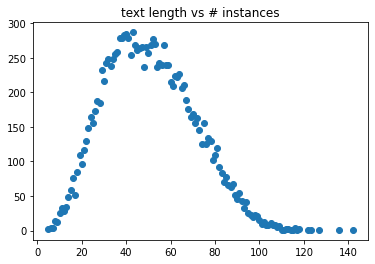}
    \caption[LJSpeech Dataset Text Length vs Number of Instances]{LJSpeech Dataset Text Length vs Number of Instances}
    \label{fig:ljspeech_text_length_vs_number_of_instance}
\end{figure}

From the graphical analysis, it's evident that the distribution of my dataset is highly random. This meets the requirement for a random dataset distribution for testing purposes. In contrast, the LJSpeech dataset exhibits a more uniform distribution, particularly in the "Text Length vs Number of Instances" comparison, which closely aligns with a normal distribution.

\subsection{Dataset SNR Analysis}
Signal-to-Noise Ratio (SNR) is a critical measure in the field of signal processing and communications. It represents the proportion between the power of a signal and the power of background noise. Higher SNR values indicate that the signal is of higher quality, as it's more distinguishable from the noise. Conversely, a lower SNR suggests that the signal quality is poor because the noise level is relatively high.

The SNR is usually measured in decibels (dB) and is calculated using the formula:

\begin{equation*}
SNR (dB) = 10 \log_{10} \left( \frac{P_{\text{signal}}}{P_{\text{noise}}} \right)
\end{equation*}
where \( P_{\text{signal}} \) is the power of the signal and \( P_{\text{noise}} \) is the power of the noise.

In audio processing, a higher SNR generally indicates clearer, more discernible sounds, while in wireless communications, a higher SNR can lead to more reliable data transfers and fewer errors. Evaluating the SNR is crucial in various applications to ensure the accuracy and quality of signal transmission or retrieval.

\begin{figure}[ht]
    \centering
    \includegraphics[width=1.0\textwidth]{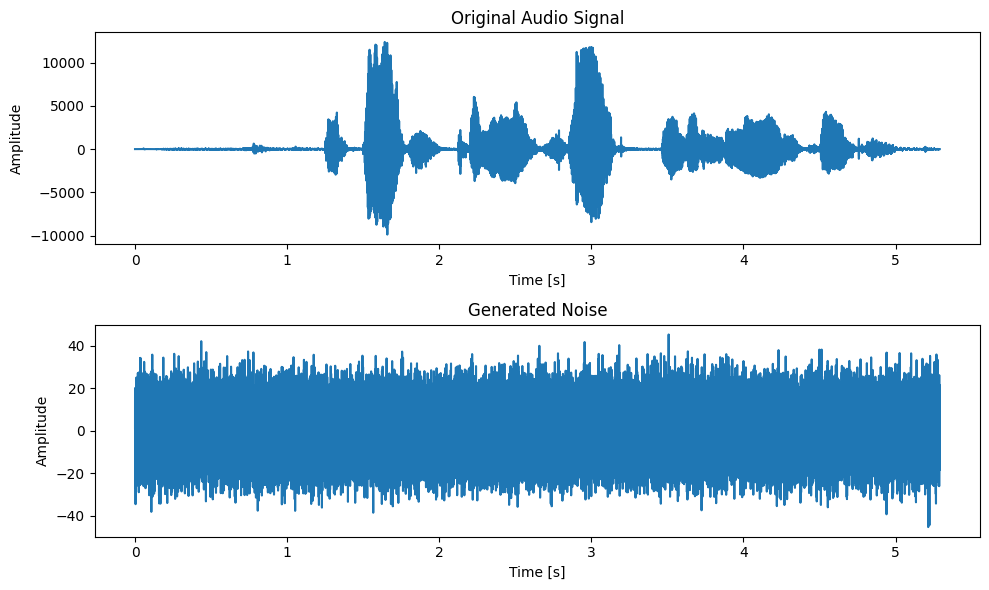}
    \caption[My Dataset SNR]{My Dataset SNR}
    \label{fig:My Dataset SNR}
\end{figure}

According to \cite{kim2008robust}, a Signal-to-Noise Ratio (SNR) less than 20 dB indicates minimal interference from noise in a dataset. In the context of my research, the average SNR was measured at 11.69 dB, suggesting that noise would not significantly influence the outcomes of my model. Thus, there's no need for additional noise processing on my dataset.

\section{Data Preprocessing}
\subsection{Data Cleaning}

A text cleaner serves as a preprocessing tool in voice and speech processing projects, ensuring that the input text data is standardized and free of anomalies. By removing or modifying inconsistent formatting, punctuations, special characters, and other potential noise factors, a text cleaner aids in creating a uniform dataset. This uniformity is crucial as models often perform better and train faster on consistent and clean data. Furthermore, by converting different textual representations of numbers, dates, or abbreviations to a consistent format, text cleaners enhance the quality of the generated speech in voice cloning or synthesis projects. In essence, a text cleaner ensures the textual data's integrity, paving the way for better model training and improved speech output quality. The text formatter used in this paper is highly inspired by the work of \cite{yang2021torchaudio}.

The text cleaning procedure is as follows:

\textbf{Case Normalization}: All letters in the text are converted to lowercase to maintain consistency and avoid duplication due to case variations.

\textbf{Time Conversion}: Time representations are converted into textual form to make them comprehensible when read aloud. For instance, "2:30pm" is translated to "two thirty p m", and "2:05" is translated to "two oh five".

\textbf{Numerical Normalization}: 
Recognizing and removing commas in numbers for a seamless readout.
Converting currency symbols followed by numbers into their full textual representation. As an example, "\$100" is translated into "one hundred dollars".
Decimal numbers are also dealt with, with "3.14" being read as "three point one four".
Ordinal numbers like "21st" are transformed into "twenty first".
Generic numbers are translated into their word form for clarity.

\textbf{Abbreviation Expansion}: Common abbreviations in the text are expanded to their full form to ensure clarity in speech output. For instance, "co" becomes "company", "gen" transforms into "general", and "mr" is read as "mister".

\textbf{Symbol Translation}: Symbols that can be verbalized are transformed into their respective word forms for clearer interpretation. An example would be the conversion of "\&" into "and".

\textbf{Auxiliary Symbol Removal}: Certain symbols that don't contribute to the semantic meaning, such as "<", ">", "(", and ")", are removed from the text.

\textbf{Whitespace Normalization}: Multiple consecutive whitespace characters, including spaces, tabs, or newline characters, are replaced with a single space. Additionally, any whitespace at the beginning or end of the text is trimmed off.

\section{Phoneme Conversion}
The use of phonemes in speech synthesis and recognition models has become a widespread practice \cite{NLPbook}. As the fundamental units of sound in language, phonemes play a pivotal role in achieving multiple key objectives. Firstly, they aid in standardization by reducing disparities across different languages and dialects, enhancing the adaptability of models. Secondly, phonemes increase the flexibility in speech synthesis, enabling the capture of richer vocal nuances, paving the way for more expressive synthesized voices \cite{NLPbook}. Furthermore, phonemes enhance the efficiency and accuracy of models, given that they directly map the relationship between text and sound, especially in languages where the relationship between letters and their pronunciation isn't consistent. Additionally, the multi-language processing capability of phonemes is another reason for their popularity, as many languages share the same phonemes, facilitating the development of multilingual models. Lastly, phonemes assist models in better generalizing when faced with unfamiliar text, as they can more effectively infer the pronunciation of unknown words \cite{NLPbook}. Therefore, the use of phonemes not only bolsters the interpretability and accuracy of models but also augments their flexibility. This is the primary reason many modern voice models opt to use phonemes as the basic unit for text processing \cite{NLPbook}.

In our research, we utilized an open-source package called Phonemizer \cite{Bernard2021} to convert our text into the International Phonetic Alphabet (IPA). The IPA is a universally accepted system devised to accurately depict and record phonemes of any human language \cite{international1999handbook}. Its primary aim is to provide a unified and unambiguous representation for all human speech sounds. As a widely accepted international standard, the IPA holds significant traction in linguistics and related fields \cite{international1999handbook}. One of its core strengths lies in its intricate detailing, allowing for the specific portrayal of almost any conceivable human speech sound, transcending the confines of a particular language or dialect. This system is not only standardized but also scalable, adapting to novel sound descriptions. Contrarily, many languages' orthographic systems don't always reflect their actual pronunciation. The IPA offers a way around these orthographic constraints, enabling clear recording of these pronunciations \cite{international1999handbook}. For instance, the sentence "I love eating apples," when transcribed in standard American English pronunciation, is represented in IPA as \textipa{/aI l\textturnv v \textprimstress itIN \textprimstress \ae plz}
/. This method of transcription provides a more precise and granular record of pronunciation, aiding a deeper understanding and description of the sonic structure of languages.

\section{Forced Alignment}
Forced Alignment is a technique employed in the realm of speech processing, primarily aiming to automatically align pronunciations within an audio with its corresponding textual transcription. In essence, this method facilitates the automatic annotation of specific starting and ending time points of pronunciations or phonemes for a given audio segment.

Here are several reasons why Forced Alignment is used:

\textbf{Data Annotation}:
During the preparation of training data for speech recognition or synthesis, there's often a need to label time-specific information for phonemes or words across vast audio data sets. Manual execution of this task is both time-consuming and labor-intensive. In contrast, Forced Alignment accomplishes this process in an automated and expedited manner.

\textbf{Speech Research}:
In linguistic or phonetic research, scholars might be intrigued by specific phonetic phenomena, such as the duration of sounds or the placement of stress. Forced Alignment offers researchers a method to swiftly attain precise alignment details for such studies.

\textbf{Pronunciation Assessment}:
Within the context of language learning applications, students' pronunciations can be evaluated by aligning them with standard pronunciations. As a result, learners receive specific feedback concerning deviations in their pronunciation.

\textbf{Multimodal Data Processing}:
When handling data that encompasses both audio and text, Forced Alignment ensures synchronization between the audio and the text components.

\textbf{Enhancing the Accuracy of Speech Applications}:
By ensuring a precise alignment between audio and text, speech synthesis and recognition systems can be trained with greater accuracy, thereby elevating their overall performance.

\subsubsection{Montreal Forced Aligner}
I utilized the Montreal Forced Aligner \cite{mcauliffe2017montreal} to perform forced alignment for some models. The Montreal Forced Aligner (MFA) stands as a testament to the advancement of open-source tools in the realm of speech processing \cite{mcauliffe2017montreal}. It offers a platform that researchers and developers can freely utilize and modify. Notably, MFA distinguishes itself with its ability to train on new and diverse data, allowing users to optimize alignment based on specific datasets \cite{mcauliffe2017montreal}. This adaptability is further highlighted by its use of triphone acoustic models to capture contextual variations in phoneme realization, a marked departure from aligners that predominantly rely on monophone acoustic models \cite{mcauliffe2017montreal}. The inclusion of speaker adaptation for acoustic features enriches its capacity by simulating differences between speakers. A significant facet of MFA's prowess is its foundation on Kaldi \cite{mcauliffe2017montreal}, a renowned open-source automatic speech recognition toolkit. This association provides a suite of benefits such as enhanced portability, the power of parallel processing, and the capability to scale to larger datasets \cite{mcauliffe2017montreal}. Adding to its versatility, MFA can accommodate multiple transcription formats, making it an adaptable tool for various speech data types \cite{mcauliffe2017montreal}. Furthermore, its in-built error detection and correction function assists users in pinpointing and rectifying deviations from transcriptions, addressing a common source of alignment errors \cite{mcauliffe2017montreal}. Evaluation metrics underscore MFA's superior performance compared to other open-source aligners with simpler architectures \cite{mcauliffe2017montreal}. Overall, the combination of these features and capabilities firmly positions the Montreal Forced Aligner as a robust and versatile tool in the ever-evolving landscape of speech-text alignment research.

To train Forced Alignment, a lexicon is also required; it provides standard phoneme pronunciations for a given set of vocabulary. Each word might have one or multiple potential phoneme pronunciations, depending on dialects, context, or other factors. Since my dataset is highly similar in format to LJSpeech and both are in English, I utilized the lexicon provided for LJSpeech by the MFA official repository. Additionally, I employed MFA's pre-trained English acoustic models to execute the forced alignment.

Figure 3.13 is one of the lines from my dataset using MFA's implementation of Forced Alignment, "The birch canoe slid on the somooth planks ", an aligned image output using the Praat tool. Praat is a computer software specifically designed for speech analysis, developed by Paul Boersma and David Weenink from the University of Amsterdam in the Netherlands \cite{Boersma2009}. It allows users to view spectrograms, pitch, and other speech parameters, and also offers capabilities for speech editing, annotation, and basic speech synthesis functions \cite{Boersma2009}.

\begin{figure}[ht]
    \centering
    \includegraphics[width=1.0\textwidth]{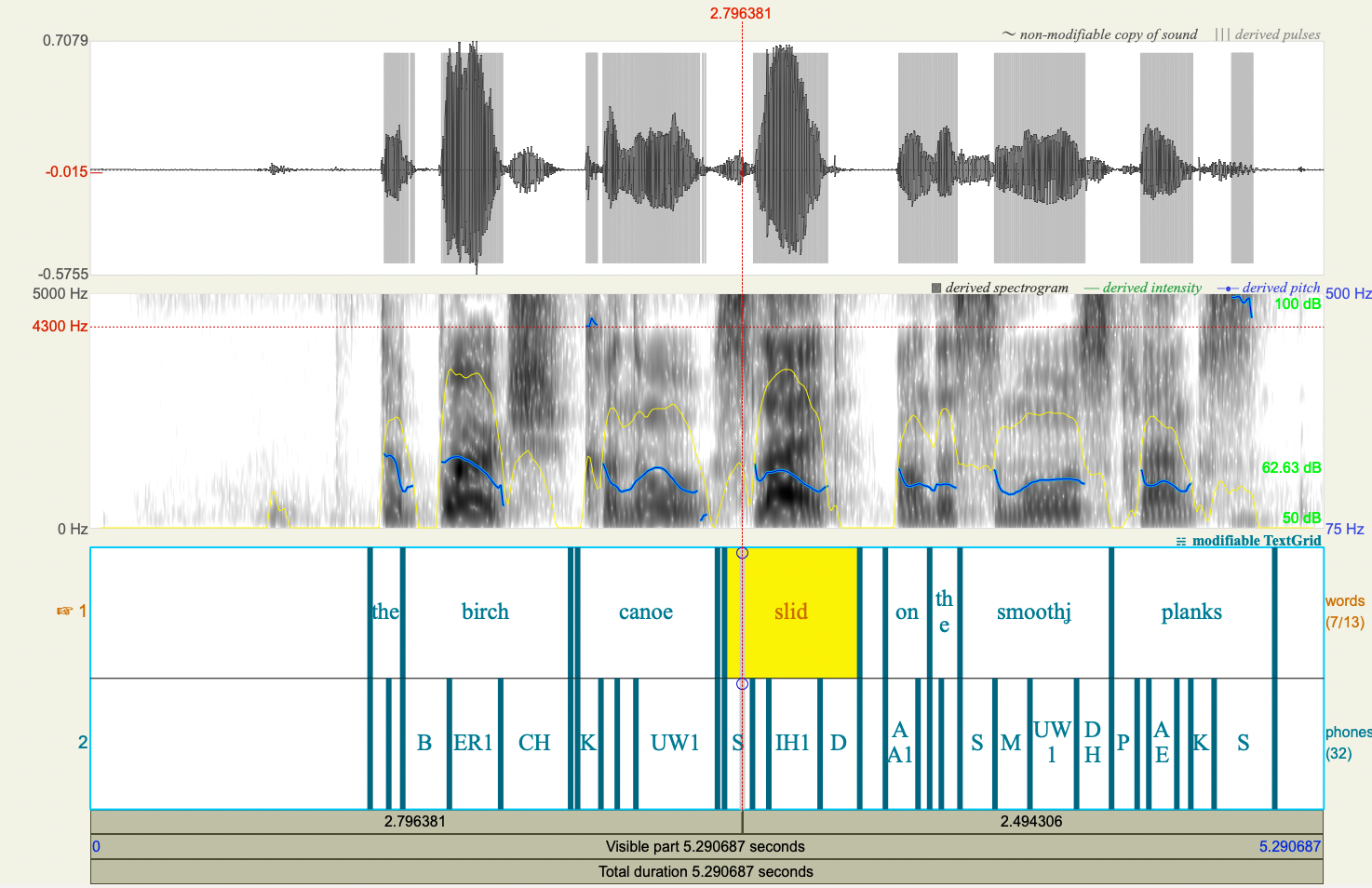}
    \caption[Montreal Forced Aligner Example]{Visual representation of the speech segment "The birch canoe slid on the smooth planks" using Praat. From top to bottom: waveform of the speech, derived spectrogram, intensity (indicated by the yellow line), pitch (represented by the blue line), and the aligned phoneme sequence.}
    \label{fig:Montreal Forced Aligner Example}
\end{figure}

The display showcases, from top to bottom, the waveform of my voice, the derived spectrogram, and the phoneme representation. Within the spectrogram, the yellow line indicates intensity, and the blue line represents pitch.

\section{Transfer Learning and Model Fine-Tuning}

\subsection{Understanding Transfer Learning}

Transfer learning is a machine learning strategy where a pre-trained model is used on a new, related task. It's the idea of overcoming the isolated learning paradigm and leveraging knowledge acquired from related tasks to improve the learning performance. In the context of deep learning for voice synthesis, models are typically pre-trained on large datasets, learning a rich understanding of the task which can then be adapted for more specific applications.

\subsection{Fine-Tuning and Its Advantages}

Fine-tuning is a form of transfer learning. Once a model has been pre-trained, fine-tuning adjusts the model parameters to the new task using the new task-specific dataset. This is especially advantageous when dealing with smaller datasets, as is the case in this research. Rather than training a model from scratch, which would require a substantial amount of data, fine-tuning allows us to leverage existing learned features and adjust them to our specific task. This can result in significant savings in both time and computational resources while potentially achieving a higher model performance than training from scratch.

\subsection{Application of Transfer Learning and Fine-Tuning in This Research}
In this study, we leverage models that have been pre-trained on the LJSpeech dataset, a large-scale English speech dataset. 
The sentences read by the speaker are sourced from public domain books available on Project Gutenberg, and they cover a variety of topics and genres, providing a rich set of different phonemes and phoneme sequences. This makes LJSpeech a versatile dataset for training text-to-speech models. The high quality and consistency of the recordings, along with the variety of sentence content, makes the LJSpeech dataset a valuable resource for training and evaluating speech synthesis systems.

These models have a comprehensive understanding of the task of text-to-speech synthesis. We then apply the strategy of fine-tuning to adapt these pre-existing deep learning text-to-speech models to our few-shot, low-resource, customized dataset. 

By using transfer learning and fine-tuning, we aim to achieve high-quality voice cloning on a smaller dataset than these models were originally trained on. It is crucial to note that all the selected models have their base learning from the same LJSpeech dataset, which ensures a level of uniformity in the base knowledge. This uniformity aids in providing a more reliable comparison of their performance post fine-tuning on our specific dataset.

This strategy not only enables a comparison of how different architectures respond to fine-tuning on a small, specific dataset, but it also contributes to our understanding of their suitability for low-resource voice cloning applications.

%% file: experiment.tex
\chapter{Experiments}

\section{Hyper-parameter Selection}
This study fine-tuned pre-trained models of Tacotron2, FastSpeech2, FastPitch, Glow-TTS, VITS, SpeechT5, and OverFlow on the LJSpeech Dataset. The training was conducted on Google Colab, where the setup included 83.5 GB of RAM and an NVIDIA A100 GPU with CUDA support. 
 In the course of the research, the primary objective was to strike a balance between producing high-quality audio outputs and maintaining computational efficiency, especially under constraints of limited resources. To guarantee a consistent and unbiased comparison amongst various models, specific key hyperparameters were standardized.

\begin{itemize}
    \item \textbf{Learning Rate}: The selected learning rate was \(1e-3\). The orginal learning rate was  \(1e-4\), however, through experimentation, it was found that the fine-tuning speed for Tacotron2 was overly slow. 
    \item \textbf{Epoch}: Initial experiments with longer epochs, 1,000 and 500, produced Tacotron2 models of subpar quality. The synthesized speech was indistinguishable and overwhelmed with noise. Furthermore, these settings resulted in prolonged training durations. Therefore, after empirical evaluation, we decided on 250 epochs.
    
    \item \textbf{Data Splitting}: Out of 49 samples, 45 were allocated for training and 5 for validation. Considering the limited size of our training dataset (45 samples), a smaller batch size was adopted. This decision ensures a sufficient number of weight updates during each epoch, allowing the model to adapt optimally to the data. Larger batch sizes might impede the frequency of weight updates, especially with such a limited dataset.

    \item \textbf{Batch Size}: Given the limited size of my training dataset, comprising only 45 samples, I opted for a smaller batch size. This decision ensures that the model undergoes sufficient weight updates in each epoch, enabling it to better learn from and adapt to the data. Using a larger batch size might restrict the frequency of weight adjustments, especially in the context of such a constrained dataset.
    
    \item \textbf{Iteration Count}: The model was trained for 3,000 iterations.
    
    \item \textbf{Warm-up Steps}: The warm-up phase was set to 200 steps. Given an iteration count of 3,000, this warm-up period offers a gradual learning rate adjustment phase, ensuring stability early in training. The model begins with a subdued learning rate, mitigating potential oscillations or instabilities. As the 200-step threshold nears, the learning rate gradually reaches its peak, adjusting as per the set strategy for the remainder of the training. This method enhances training stability and final model performance.
    
    \item \textbf{Weight Decay}: A decay rate of 0.01 was chosen for L2 regularization. Given the small dataset size, a higher penalty is applied to prevent model over-complication and overfitting.
    
    \item \textbf{Precision}: We adopted mixed-precision training using fp16 to strike a balance between computation efficiency and training stability.
    
    \item \textbf{Vocoder}: The choice for the vocoder was HiFi-GAN, a model optimized for high-fidelity speech synthesis and based on GANs. Given its commendable audio quality and considering the time costs, we opted not to train a vocoder from scratch for this research.
\end{itemize}

\begin{table}[h]
\centering
\begin{tabular}{|l|l|}
\hline
\textbf{Hyper-parameter} & \textbf{Value/Description} \\
\hline
Learning Rate & \(1e-3\) \\
\hline
Epoch & 250 \\
\hline
Data Splitting & 45 training samples, 5 validation samples \\
\hline
Batch Size & 4 \\
\hline
Warm-up Steps & 200 \\
\hline
Weight Decay (L2 regularization) & 0.01 \\
\hline
Precision & Mixed-precision (fp16) \\
\hline
Vocoder & HiFi-GAN \\
\hline
\end{tabular}
\caption{Hyper-parameters summary}
\label{tab:hyperparameters}
\end{table}

\section{Model Performance and Evaluation}

\subsection{Training and Validation Losses}

The training and validation losses serve as primary indicators of a model's performance throughout the learning process. As the model iteratively refines its parameters in response to the data, these losses provide insights into its progression, convergence behavior, and generalization capability. Training loss assesses the model's fit to the data it directly learns from, while validation loss offers a perspective on its performance with unseen data. A comparative examination of these losses aids in diagnosing potential issues such as overfitting, underfitting, or other convergence anomalies. In this section, we delve into the trajectory of these losses during the training of the VITS model and discuss their implications for model evaluation.

\subsubsection{Tacotron2 Losses}

From the visual representations of Tacotron2's training and validation loss, it's evident that the model's loss metric performs favorably. The loss consistently decreases, reaching an approximate value of 0.1 around the 2,000-step mark. This observation suggests that further training may be unnecessary beyond this point. The loss drops to a point and then smoothes out, this may mean that the model has reached convergence, i.e. the model may not benefit from further training.

\begin{figure}[ht]
\centering
\includegraphics[width=0.5\textwidth]{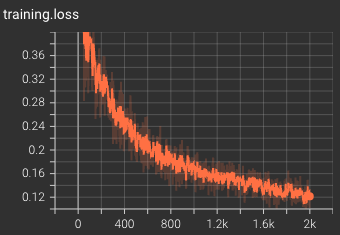}
\caption[Tacotron2 Training Loss]{This figure displays the progression of Tacotron2's training loss over time, highlighting its consistent descent towards a value close to 0.1 by the 2,000-step mark.}
\label{fig:tacotron2_training_loss}
\end{figure}

\begin{figure}[ht]
\centering
\includegraphics[width=0.5\textwidth]{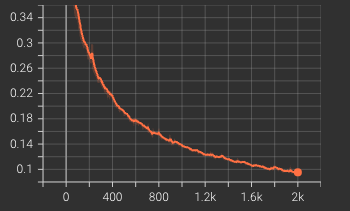}
\caption[Tacotron2 Validation Loss]{This figure illustrates the trajectory of Tacotron2's validation loss throughout its training duration, underscoring the model's ability to generalize well on unseen data.}
\label{fig:tacotron2_val_loss}
\end{figure}

The model's training and validation losses both exhibit consistent reductions. This indicates that the model's architecture is well-suited for transfer learning, especially given the limited data at hand. The model's success in fine-tuning from its pre-trained state implies that its inherent design and initialization methods are conducive to transfer learning. In situations with limited data, there's an increased risk of model overfitting. However, the steady decrease in validation loss suggests that this model effectively mitigates the risk of overfitting.

\subsubsection{FastSpeech2 Losses}

\begin{figure}[ht]
\centering
\includegraphics[width=1.0\textwidth]{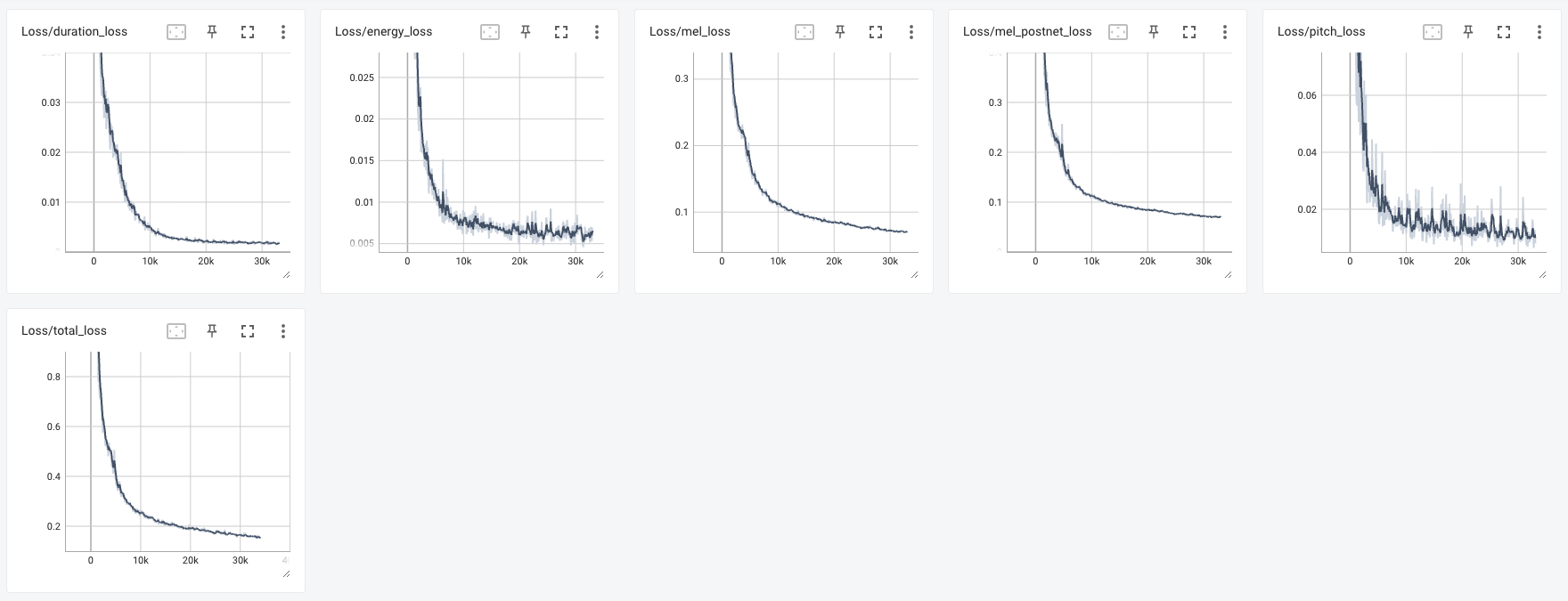}
\caption[FastSpeech2 Training Loss]{This figure illustrates the training loss trajectory for the FastSpeech2 model. The consistent decline indicates the model's progressive learning and adaptation to the training data over the iterations.}
\label{fig:fastspeech2_trianl_loss}
\end{figure}

\begin{figure}[ht]
\centering
\includegraphics[width=1.0\textwidth]{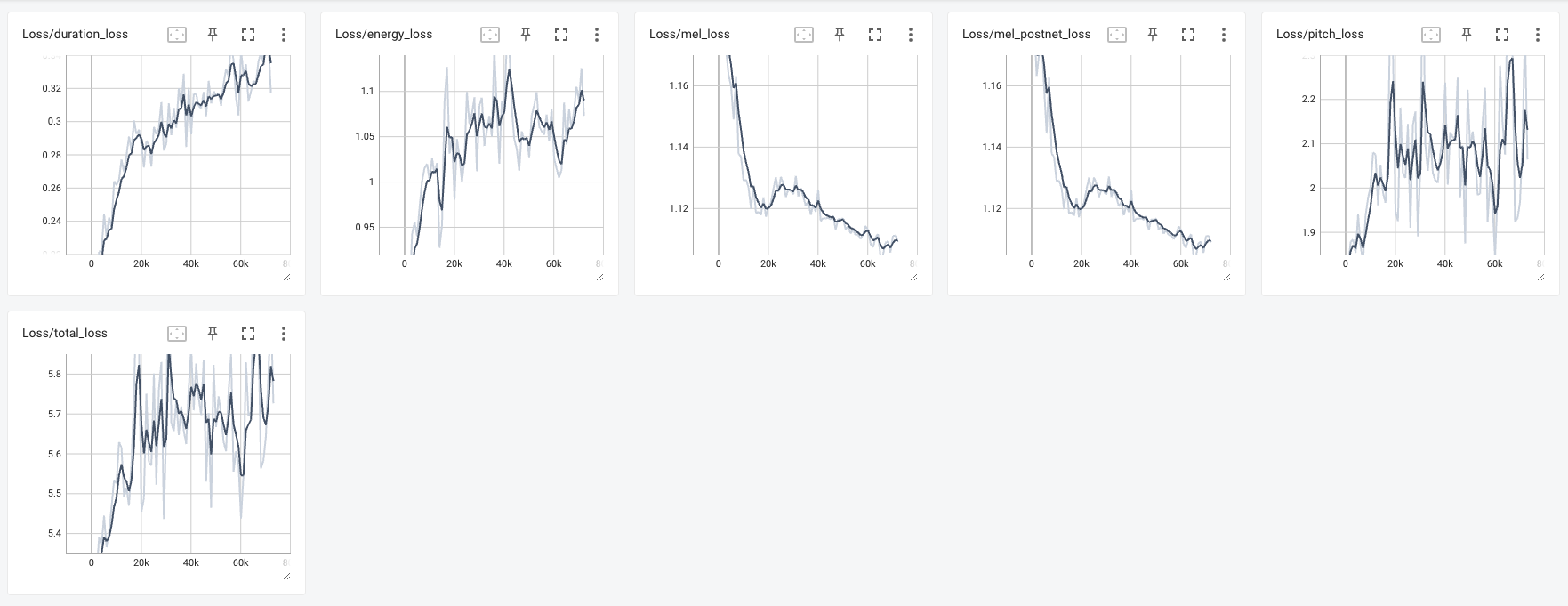}
\caption[FastSpeech2 Validation Loss]{This figure showcases the validation loss trajectory for the FastSpeech2 model. While initial decline is observed, the subsequent increase within a certain range suggests potential overfitting or challenges in generalizing to unseen data.}
\label{fig:fastspeech2_val_loss}
\end{figure}

FastSpeech2 incorporates multiple loss functions, ensuring the model's optimization across various aspects. Below is a brief explanation of these losses:

\begin{itemize}
\item \textbf{Duration loss}: This loss accounts for the discrepancy between the model's predicted duration and the actual duration. FastSpeech2 uses this loss to correctly align the input text with the output spectrogram.

\item \textbf{Energy loss}: This is related to the energy (or loudness) of the sound. The model aims to predict the energy for each frame and computes the loss against the actual energy values.

\item \textbf{Mel loss}: This loss measures the difference between the model's predicted Mel-scaled spectrogram and the target spectrogram. It ensures that the generated speech content aligns with the intended content.

\item \textbf{Mel postnet loss}: FastSpeech2 employs a post-processing network (postnet) to further refine the produced spectrogram. This loss measures the variance between the spectrogram outputted by the postnet and the target spectrogram.

\item \textbf{Pitch loss}: Pertaining to pitch prediction, the model seeks to predict the pitch for each frame and calculates the loss against actual pitch values.

\item \textbf{Total loss}: This represents the summation or combination of all the aforementioned losses. Minimizing this composite loss during training ensures the model's holistic optimization across all relevant aspects.
\end{itemize}

Despite the continuous decline and eventual convergence of the Training Loss to a minimum value, the Validation Loss has been observed to increase within a confined range. A continuous decline in the loss on the training set, coupled with a rise in the loss on the test (or validation) set, typically signals overfitting. Overfitting can stem from an overly complex model that memorizes the training data instead of discerning useful patterns from it. Such a scenario hampers performance on the test set as the model fails to generalize to unfamiliar data, may represent poor learning for the few-shots dataset.

\subsubsection{FastPitch Losses}

FastPitch, being a sophisticated neural network model, incorporates multiple loss components to ensure a comprehensive training regime that captures various aspects of speech synthesis. Each of these losses targets specific features of the voice generation process.

\begin{figure}[ht]
\centering
\includegraphics[width=1.0\textwidth]{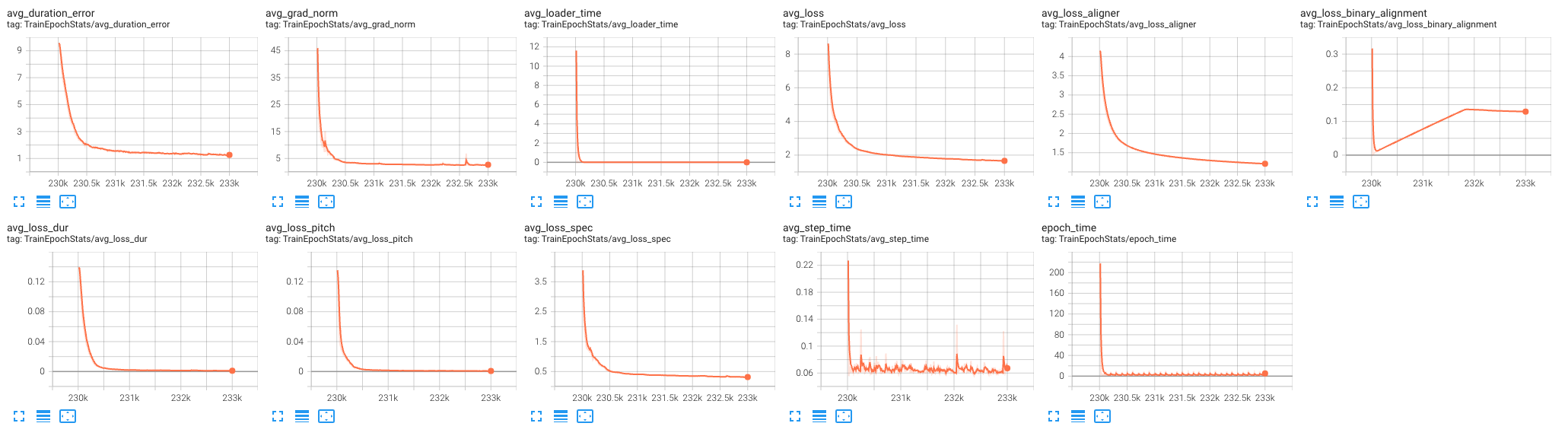}
\caption[FastPitch Training Loss]{Trajectory of training loss for the FastPitch model. A steady decrease demonstrates the model's continual learning and adaptation to the training dataset throughout the epochs.}
\label{fig:fastPitch_train_loss}
\end{figure}

\begin{figure}[ht]
\centering
\includegraphics[width=1.0\textwidth]{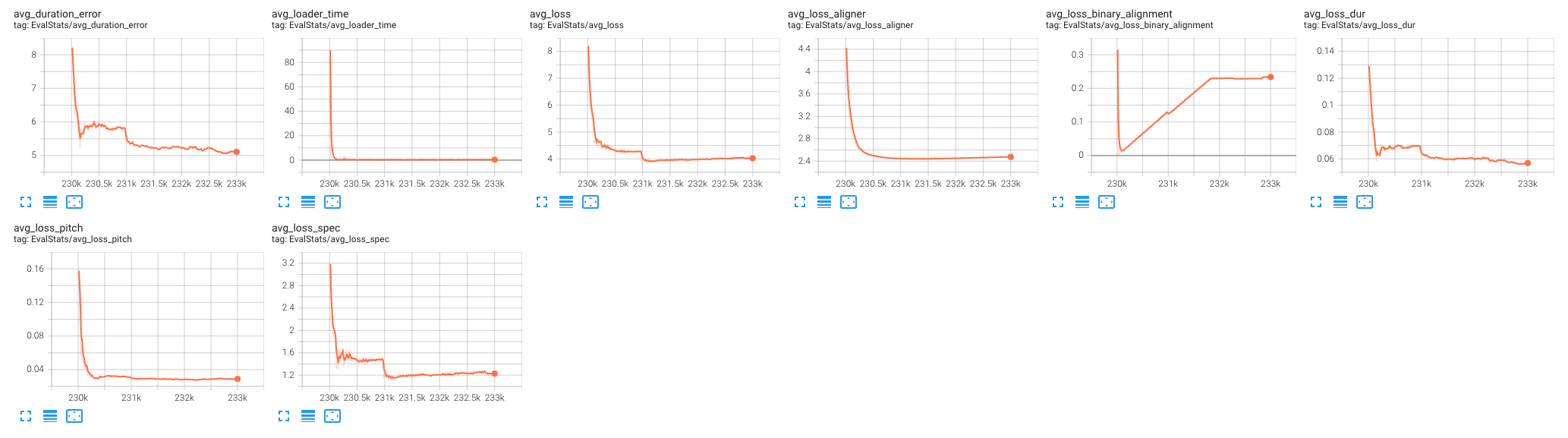}
\caption[FastPitch Validation Loss]{Validation loss progression for the FastPitch model. An initial descent followed by a plateau hints towards model stabilization. Yet, sudden spikes, especially in avg loss binary alignment, suggest potential overfitting or nuances in the data that might be challenging for the model to generalize.}
\label{fig:fastpitch_val_loss}
\end{figure}

The principal losses associated with FastPitch include:

\begin{itemize}
\item \textbf{avg duration error}: This represents the average discrepancy between the predicted speech duration by the model and the actual duration, ensuring that the generated voice aligns temporally with the original text.

\item \textbf{avg loader time}: Not strictly a loss, this metric relates to the average time taken to load training samples, indicative of data throughput efficiency.

\item \textbf{avg loss}: Likely an amalgamation or average of all other loss components, acting as the primary objective to minimize during training.

\item \textbf{avg loss aligner}: This loss pertains to the alignment between text and voice, ensuring synchronization over time.

\item \textbf{avg loss binary alignment}: A finer alignment metric between text and voice, likely capturing more granular alignment details.

\item \textbf{avg loss dur}: Analogous to avg\_duration\_error, it focuses on the alignment of voice duration.

\item \textbf{avg loss pitch}: Relates to the error between the model's predicted pitch and the actual pitch. Pitch is pivotal in both music and voice, ensuring the produced voice matches the original recording's tonality.

\item \textbf{avg loss spec}: Assesses the discrepancy between the produced and target spectrograms, guaranteeing that the synthesized voice content matches the target.

\item \textbf{avg grad norm}: In training deep learning models, the phenomenon of gradient explosion and vanishing can be encountered. On the other hand, minuscule gradients might halt the training progress, leading to gradient vanishing. To combat frequent occurrences of gradient explosions, gradient clipping is applied, ensuring the gradient values don't surpass a set threshold. Observations of gradient behaviors are pivotal as they can suggest necessary adjustments to the learning rate. Furthermore, regularization methods, such as weight decay or dropout, influence the gradient's magnitude. Hence, keeping an eye on the avg grad norm becomes essential to understand how these regularization techniques are affecting the training process.
\end{itemize}

The avg loss binary alignment, either during training or validation, plummet suddenly and then converge, is intriguing. Particularly, the abrupt drop followed by a surge in avg loss binary alignment suggests that this module, might be more challenging to minimize than other losses. As the model increasingly tailors itself to the training data, it's plausible that at some juncture, it over-optimizes for a specific loss, neglecting others in the process.

In conclusion, understanding and monitoring these loss components and metrics is vital for an effective training regime, aiding in model diagnostics, improvements, and achieving robust voice synthesis.

\subsubsection{Glow-TTS Losses}

\begin{figure}[ht]
\centering
\includegraphics[width=1.0\textwidth]{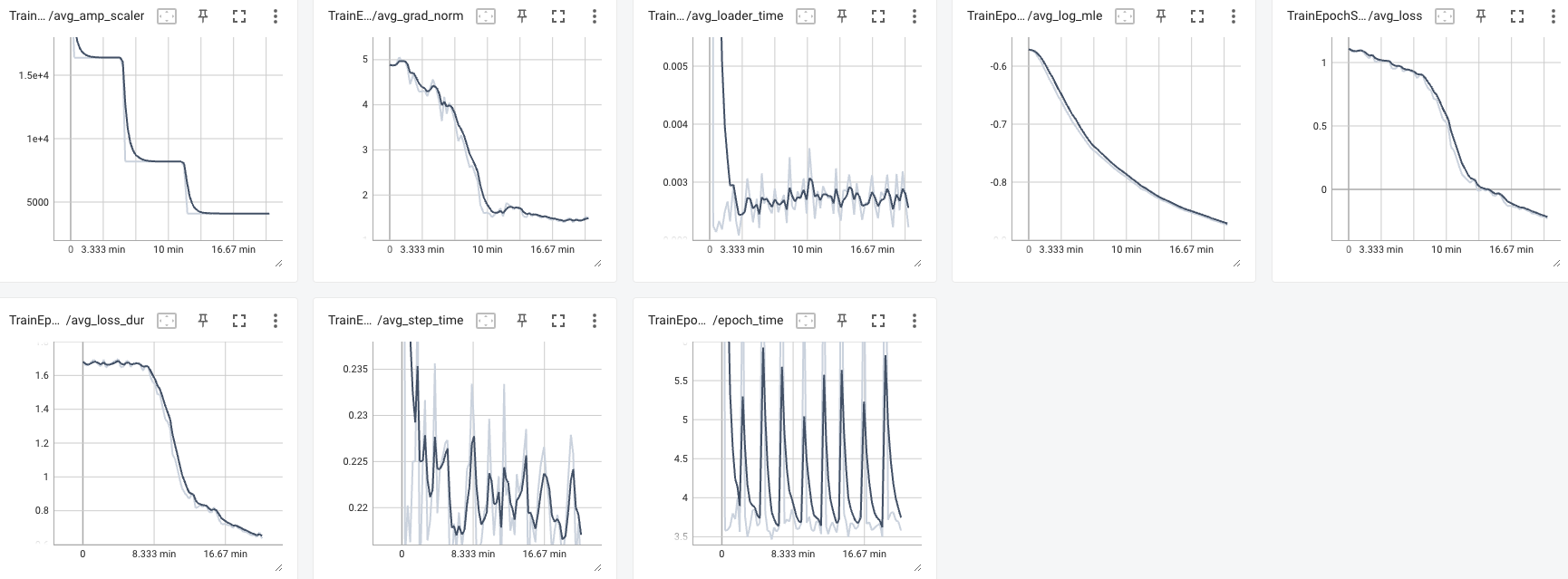}
\caption[Glow-TTS Training Loss]{Training Average Loss trajectory over iterations. The plot depicts a steady decline, indicating effective learning and adaptation to the training data.}
\label{fig:glowtts_train_loss}
\end{figure}

\begin{figure}[ht]
\centering
\includegraphics[width=1.0\textwidth]{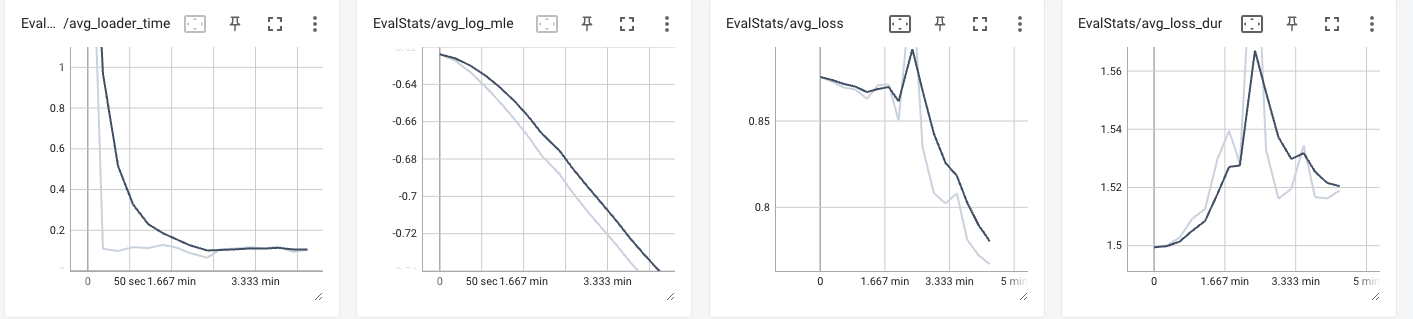}
\caption[Glow-TTS Validation Loss]{Test Set Average Loss over iterations. Noticeable spikes suggest potential distribution mismatches between training and test sets.}
\label{fig:glowtts_val_loss}
\end{figure}

The avg amp scalar is associated with mixed precision training of the model to ensure numerical stability. In mixed precision training, a combination of single and half precision is utilized to expedite training and reduce memory consumption. To circumvent potential numerical stability issues introduced by half-precision computations, particularly during weight updates, a dynamically adjusted amp scalar (automatic mixed precision scalar) is introduced to scale the model's loss. The avg amp scalar represents the average value of this scaling factor throughout the training process, offering insights into the model's numerical stability and behavior during mixed precision training.

Avg loss dur refers to the average loss associated with duration predictions in speech synthesis models. Specifically, it measures the discrepancy between the model's predicted durations for phonemes or words and their actual durations in the training data. This loss is crucial because the rhythmic flow and pace of the synthesized speech depend on accurate duration predictions. Monitoring avg loss dur helps in ensuring that the synthesized speech aligns temporally with the original text and maintains a natural rhythm.

In our observations from the plots, the average loss during training demonstrates a gradual descent. However, conspicuous spikes or "surges" are evident in the avg loss dur and avg loss on the test set. A potential explanation for this disparity could be the diminutive size of our test dataset, comprising merely five samples, which could lead to a considerable distribution gap compared to the training data, consequently affecting the inference outcomes adversely. Another plausible cause could be overfitting, where the model might have excessively adapted to the training data, compromising its generalization capabilities. Furthermore, the oscillations in loss might also be attributed to the inherent variability introduced by the mini-batch training approach.

\subsubsection{OverFlow TTS Losses}

In the OverFlow model, the log-likelihood loss is employed as the primary evaluation metric. Consequently, negative loss values are expected, as they represent the negative logarithm of the probability.
\begin{figure}[ht]
\centering
\includegraphics[width=1.0\textwidth]{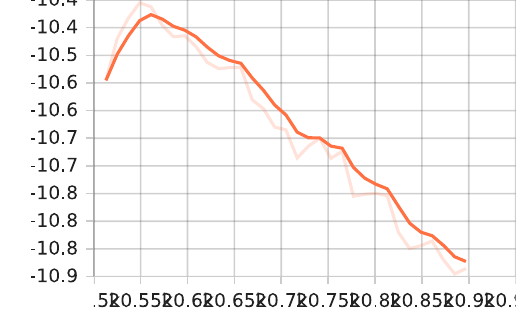}
\caption[OverFlow Training Loss]{Training loss for the OverFlow model, initially exhibiting an increase followed by a steady decline over the training epochs}
\label{fig:overflow_train_loss}
\end{figure}

\begin{figure}[ht]
\centering
\includegraphics[width=1.0\textwidth]{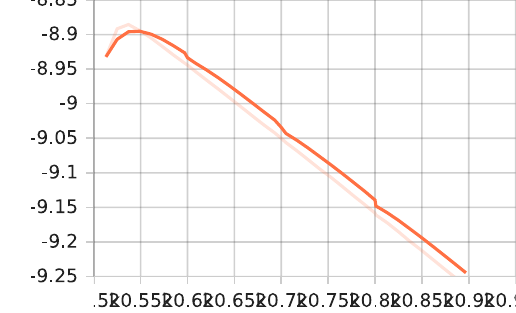}
\caption[OverFlow Validation Loss]{alidation loss for the OverFlow model, initially showing an upward trend before consistently decreasing as the model is validated across epochs.}
\label{fig:overflow_val_loss}
\end{figure}

Given that the loss is negative, it signifies that the gap between the predicted probability distribution of the model and the actual distribution is decreasing. A lesser negative log-likelihood value (closer to 0) indicates a more accurate prediction by the model. The transition of the loss from -8.95 to -9.25 suggests increased uncertainty in the model's predictions. In the context of log-likelihood loss, a smaller negative value represents a lower prediction probability, possibly indicating a decline in the model's confidence for certain samples. This phenomenon suggests that the OverFlow model may not be well-suited for few-shot, low-resource transfer learning scenarios.

\subsubsection{VITS Losses}

\begin{figure}[ht]
\centering
\includegraphics[width=0.7\textwidth]{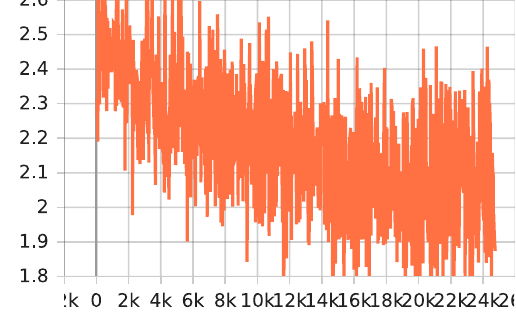}
\caption[VITS Discriminator Total Loss]{Trajectory of training loss for the VITS model. A steady decrease demonstrates the model's continual learning and adaptation to the training dataset throughout the epochs.}
\label{fig:VITS_train_loss}
\end{figure}

\begin{figure}[ht]
\centering
\includegraphics[width=0.7\textwidth]{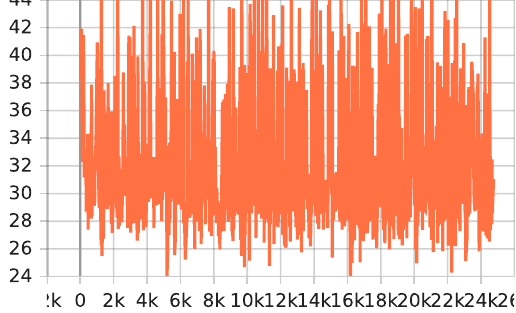}
\caption[VITS Generator Total Loss]{Trajectory of training loss for the VITS model. A steady decrease demonstrates the model's continual learning and adaptation to the training dataset throughout the epochs.}
\label{fig:VITS_losd}
\end{figure}

The definition of final loss for training the conditional VAE (Variational Autoencoder) in VITS \cite{kim2021conditional} is expressed as:

\begin{equation*}
L_{\text{vae}} = L_{\text{recon}} + L_{\text{kl}} + L_{\text{dur}} + L_{\text{adv}}(G) + L_{\text{fm}}(G)
\end{equation*}
Where:
\begin{itemize}
\item \( L_{\text{recon}} \) is the reconstruction loss.
\item \( L_{\text{kl}} \) is the KL divergence loss.
\item \( L_{\text{dur}} \) is the duration loss.
\item \( L_{\text{adv}}(G) \) is the adversarial loss for the generator.
\item \( L_{\text{fm}}(G) \) is the feature-matching loss for the generator. \\
\end{itemize}

The discriminator loss is given by:
\begin{equation*}
L_{\text{adv}}(D) = \mathbb{E}_{(y,z)}\left[\left(D(y)-1\right)^2\right] + \left(D(G(z))\right)^2
\end{equation*}

This loss function is used to train the discriminator to distinguish between the real waveform y and the generated waveform G(z). \\

The generator losses is given by:
\begin{itemize}

\item Adversarial Loss for the Generator:
\begin{equation*}
L_{\text{adv}}(G) = \mathbb{E}_{z}\left[\left(D(G(z))-1\right)^2\right]
\end{equation*}

This loss function encourages the generator to produce waveforms that the discriminator cannot easily distinguish from real waveforms.

 \item Feature-Matching Loss:
\begin{equation*}
L_{\text{fm}}(G) = \mathbb{E}{(y,z)}\left[\frac{1}{\sum{l=1}^{T}||\mathbf{D}_l(y) - \mathbf{D}_l(G(z))||_1}\right]
\end{equation*}

This loss ensures that the generated output matches the features of the real data, where T denotes the total number of layers in the discriminator, and $\mathbf{D}_l$ represents the l-th layer's output.

\end{itemize}

In the context of the VITS model, the terms \(d_g\) and \(d_r\) denote the generator loss and the discriminator loss respectively. Throughout the training process, rapid fluctuations in the loss values can be observed. According to feedback from a discussion on this GitHub issue, such behavior is typical during the training of the VITS model. Given the non-intuitive nature of these loss values, it's challenging to directly infer the model's performance or its adaptability to the dataset based on them alone.

\subsubsection{SpeechT5 Losses}

\begin{figure}[ht]
\centering
\includegraphics[width=0.7\textwidth]{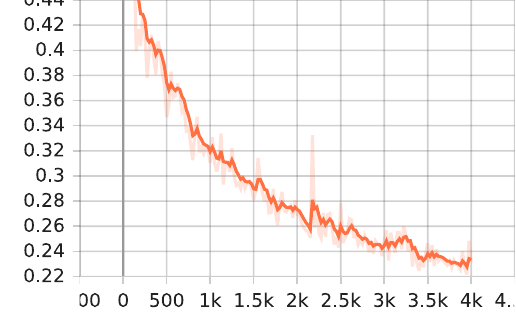}
\caption[SpeechT5 Training Loss]{Trajectory of the training loss for the SpeechT5 model. The loss curve demonstrates the model's learning progress over epochs, highlighting moments of steady decrease and occasional fluctuations}
\label{fig:SpeechT5_train_loss}
\end{figure}

\begin{figure}[ht]
\centering
\includegraphics[width=0.7\textwidth]{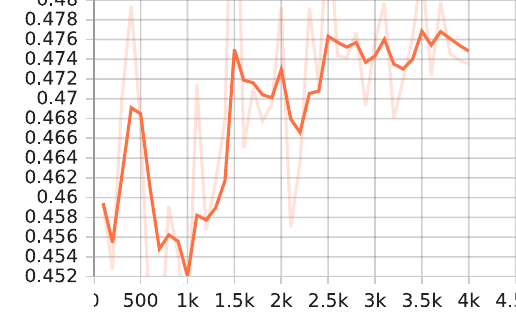}
\caption[SpeechT5 Validation Loss]{Evolution of the validation loss for the SpeechT5 model. This graph provides insights into the model's generalization capabilities across different epochs, emphasizing points of stabilization and variance.}
\label{fig:SpeechT5_evl_loss}
\end{figure}

During the training process of the model, we observed that the loss steadily decreased and eventually converged to a specific value. Although there were sawtooth-like fluctuations during training, such oscillations are within expected norms. Several factors might contribute to these fluctuations: on one hand, a relatively high learning rate and a small batch size could cause the model to jump during optimization; on the other hand, an excessively high momentum when using the Adam optimizer might lead to similar outcomes. Moreover, other factors could also impact the stability of the model's training.

In the testing phase, the model's loss exhibited significant oscillations. We hypothesize that this is directly related to the small size of our test set. A smaller test set might cause the model's evaluation to be influenced by individual or a few samples, leading to substantial changes in the loss value. To further stabilize the model's testing performance in future work, we might need to increase the size of the testing data or make other related adjustments.

%% file: Result.tex
\chapter{Result}

\section{Alignment Chart and Mel-Spectrogram Chart}

In the world of sequence-to-sequence deep learning models, an alignment chart serves as a critical visualization tool, providing a pictorial representation of the correspondence between input and output sequences. Such charts are of paramount importance in applications like machine translation or speech synthesis. By closely observing these charts, one can gain insights into how the model correlates with the input sequence during the generation of each output segment. This visualization is not just an academic exercise; it provides researchers with a valuable perspective on the model's internal processes, simultaneously facilitating the identification and resolution of potential alignment glitches. Predominantly, alignment charts have become synonymous with attention mechanisms, given that the latter’s weights transparently showcase the alignment dynamics. 

FastSpeech2, it stands apart as a non-autoregressive speech synthesis model \cite{ren2020fastspeech}. Unlike Tacotron-like models underpinned by attention, FastSpeech2 abstains from generating a conventional alignment chart to represent text-audio alignments \cite{ren2020fastspeech}. Its design is engineered to mitigate potential alignment issues inherent to attention mechanisms, especially in the initial training phases. Achieving this requires the model to utilize pre-computed durations for text-to-audio, rather than dynamically determining these via attention mechanisms \cite{ren2020fastspeech}. These durations are generally computed beforehand by models like Tacotron2 and fed into FastSpeech2 \cite{ren2020fastspeech}. Thus, while FastSpeech2 doesn’t intrinsically generate an alignment chart, the Tacotron2 model used to furnish durations can provide corresponding attention weights that can serve as a visual alignment guide. Beyond this, FastSpeech2 predicts mel-spectrogram charts and other features, including f0 (fundamental frequency) and energy \cite{ren2020fastspeech}, enhancing the naturalness and intonation of the produced voice.

\begin{figure}[ht]
\centering
\begin{minipage}{.45\textwidth}
  \centering
  \includegraphics[width=0.9\linewidth]{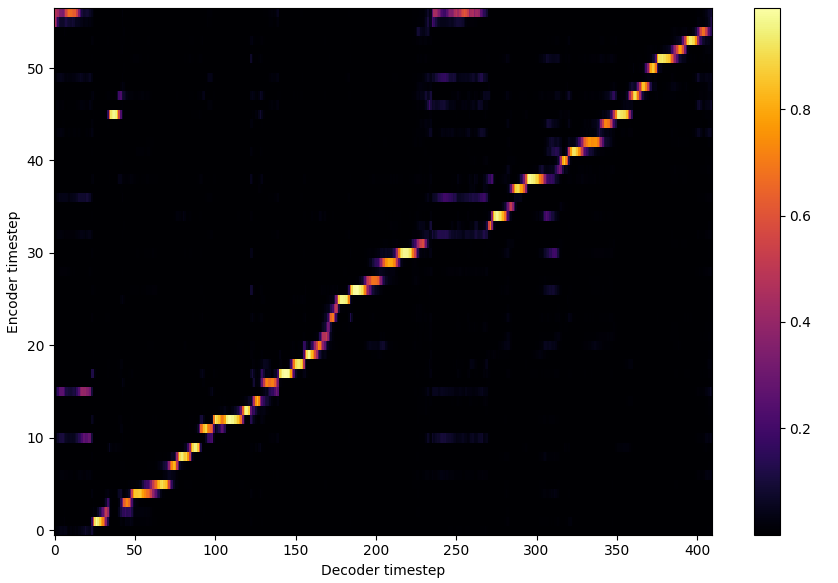}
  \caption[Tacotron2 Alignment]{Alignment chart of Tacotron2, showing the correspondence between input text and acoustic features.}
  \label{fig:tacotron2_alignment}
\end{minipage}%
\hfill
\begin{minipage}{.45\textwidth}
  \centering
  \includegraphics[width=0.9\linewidth]{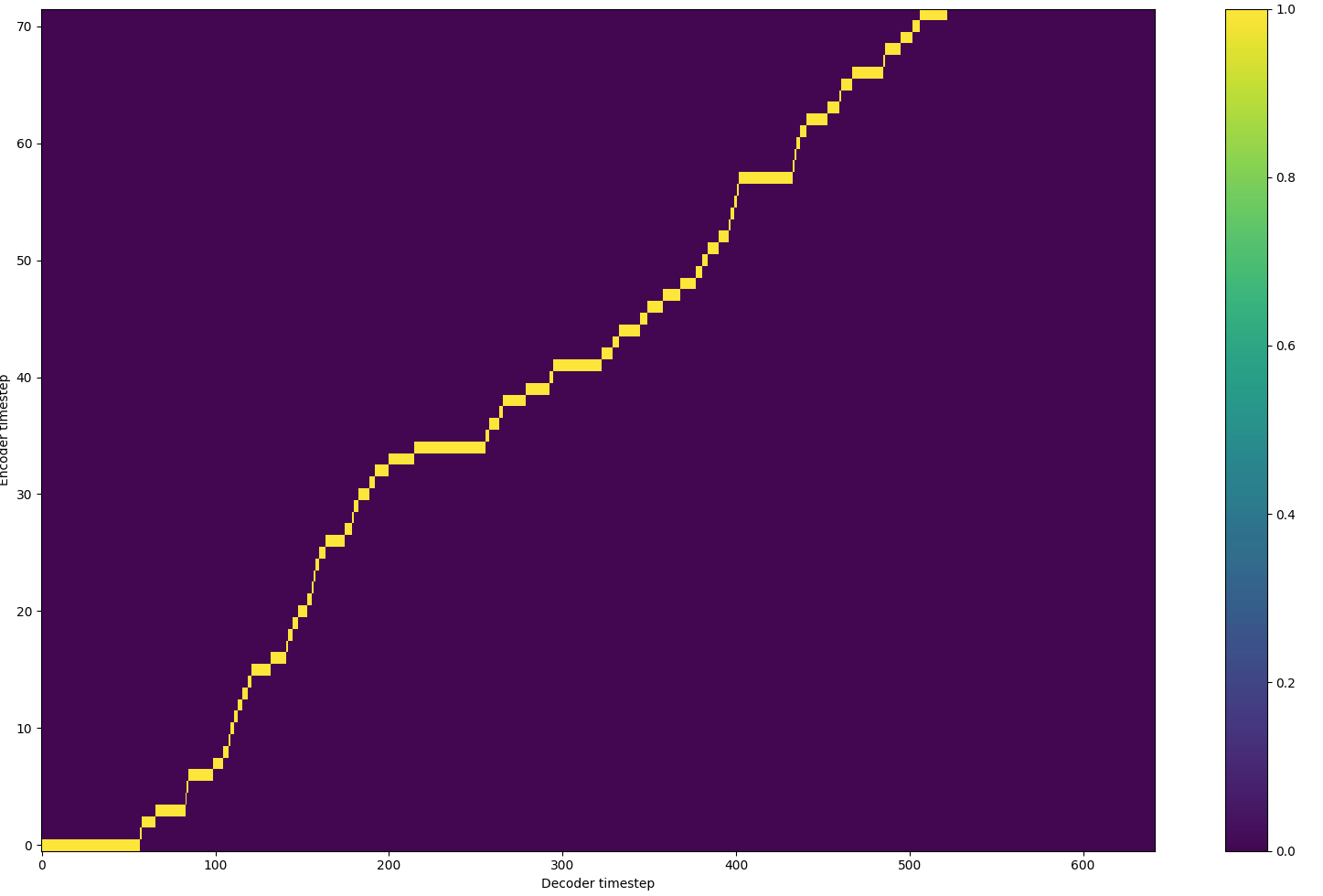}
  \caption[Glow-TTS Alignment]{Alignment visualization in Glow-TTS, demonstrating the model's ability to align text with speech signals.}
  \label{fig:glow-tts_alignment}
\end{minipage}
\end{figure}

\begin{figure}[ht]
\centering
\begin{minipage}{.45\textwidth}
  \centering
  \includegraphics[width=0.9\linewidth]{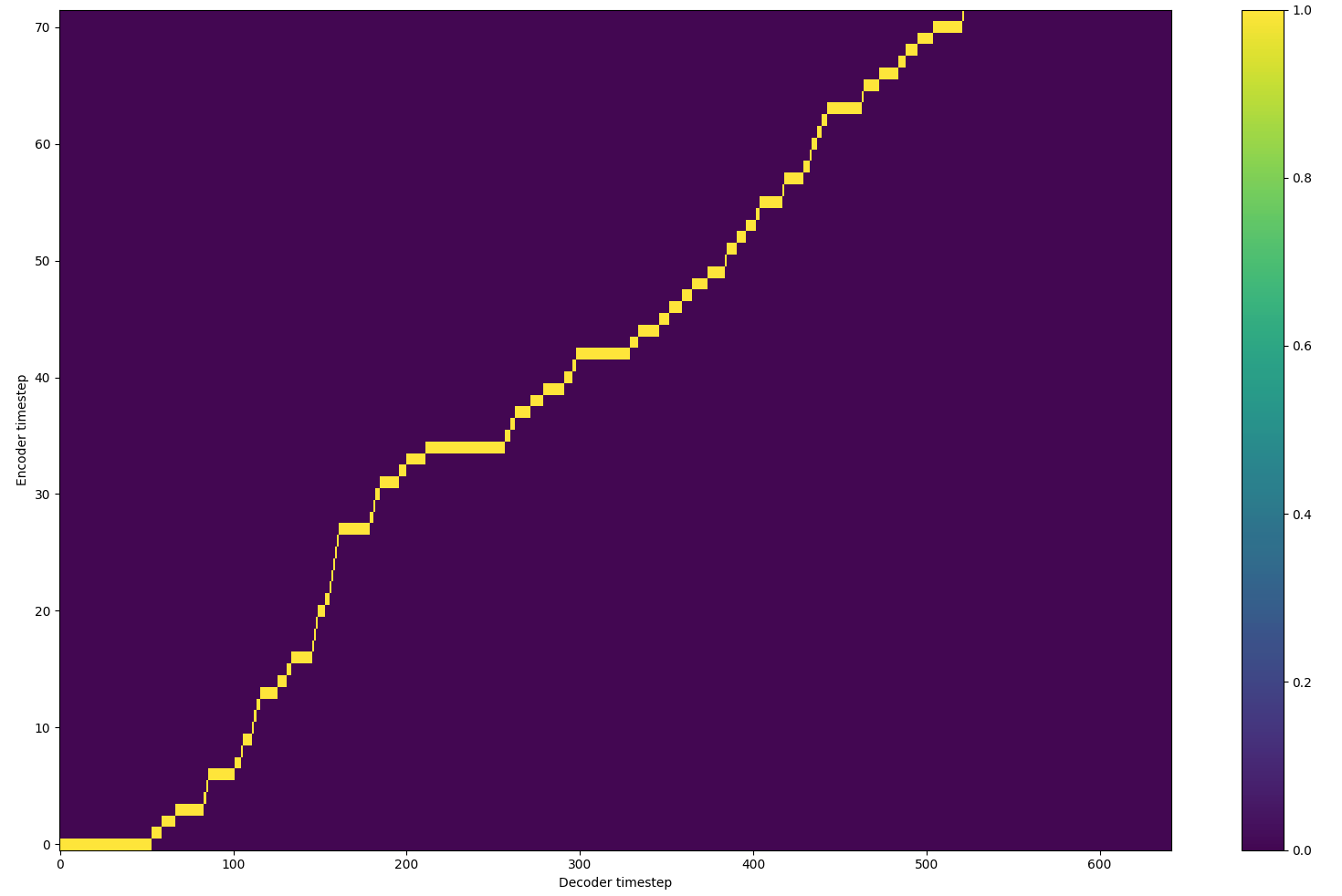}
  \caption[FastPitch Alignment]{Alignment chart in FastPitch, illustrating the traditional attention alignment between text and generated speech features.}
  \label{fig:fastPitch_alignment}
\end{minipage}
\hfill
\begin{minipage}{.45\textwidth}
  \centering
  \includegraphics[width=0.9\linewidth]{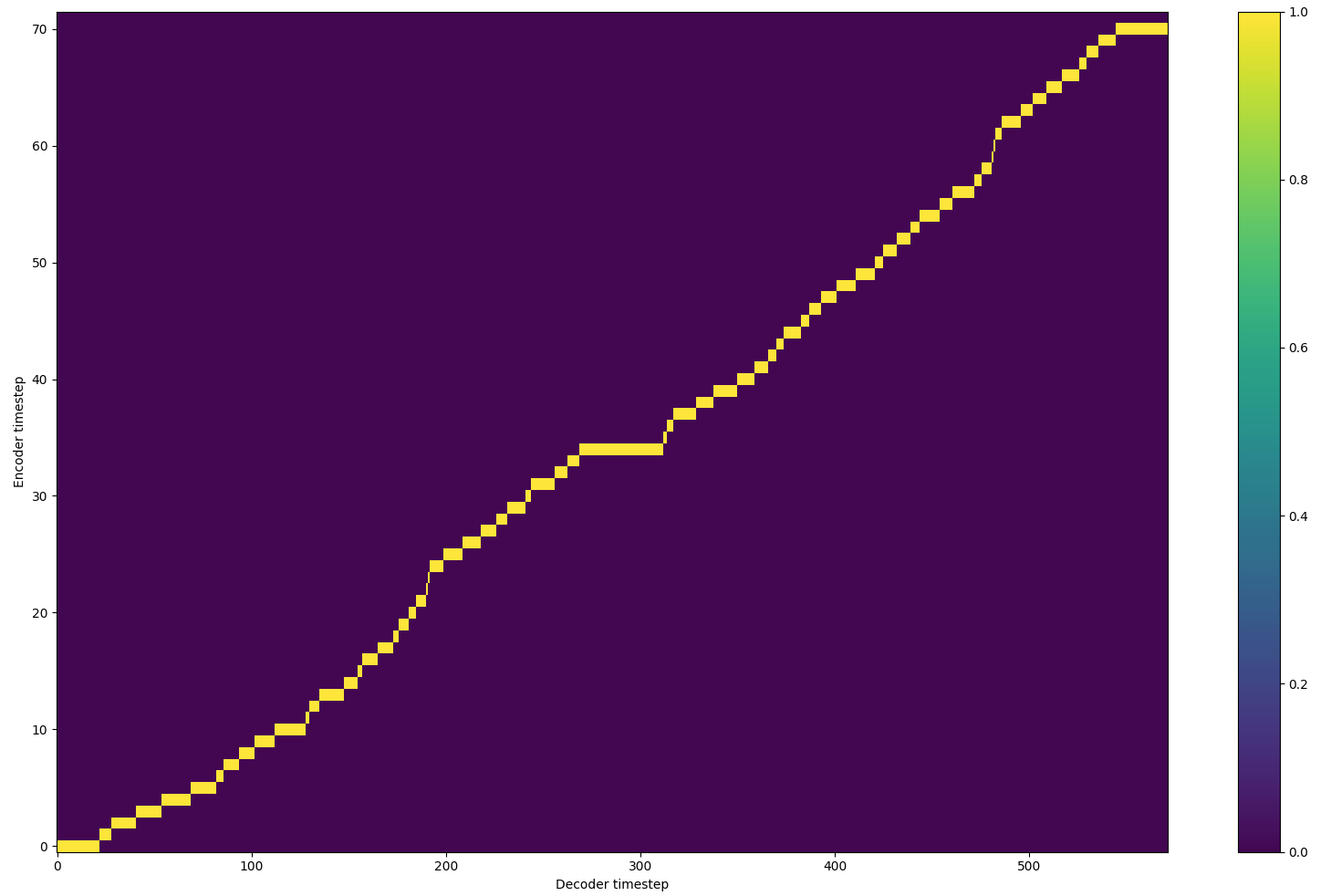}
  \caption[FastPitch Alignment Head]{The specialized “alignment head” in FastPitch, providing a clearer and more explicit view of text-to-speech alignment.}
  \label{fig:fastPitch_alignment_head}
\end{minipage}%
\end{figure}

\begin{figure}[ht]
\centering
\begin{minipage}{.45\textwidth}
  \centering
  \includegraphics[width=0.9\linewidth]{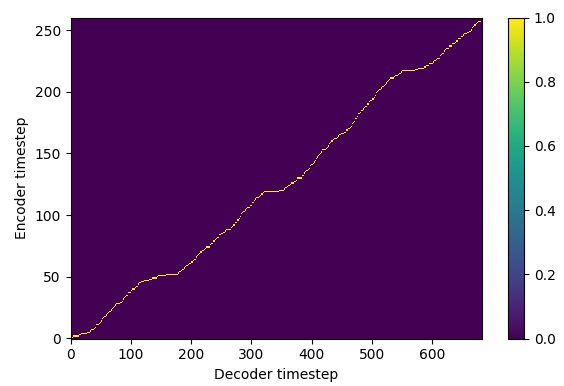}
  \caption[VITS Alignment]{Alignment visualization in VITS, showcasing the model's self-attention patterns in aligning sequences.}
  \label{fig:VITS_Alignment}
\end{minipage}
\hfill
\begin{minipage}{.45\textwidth}
  \centering
  \includegraphics[width=0.9\linewidth]{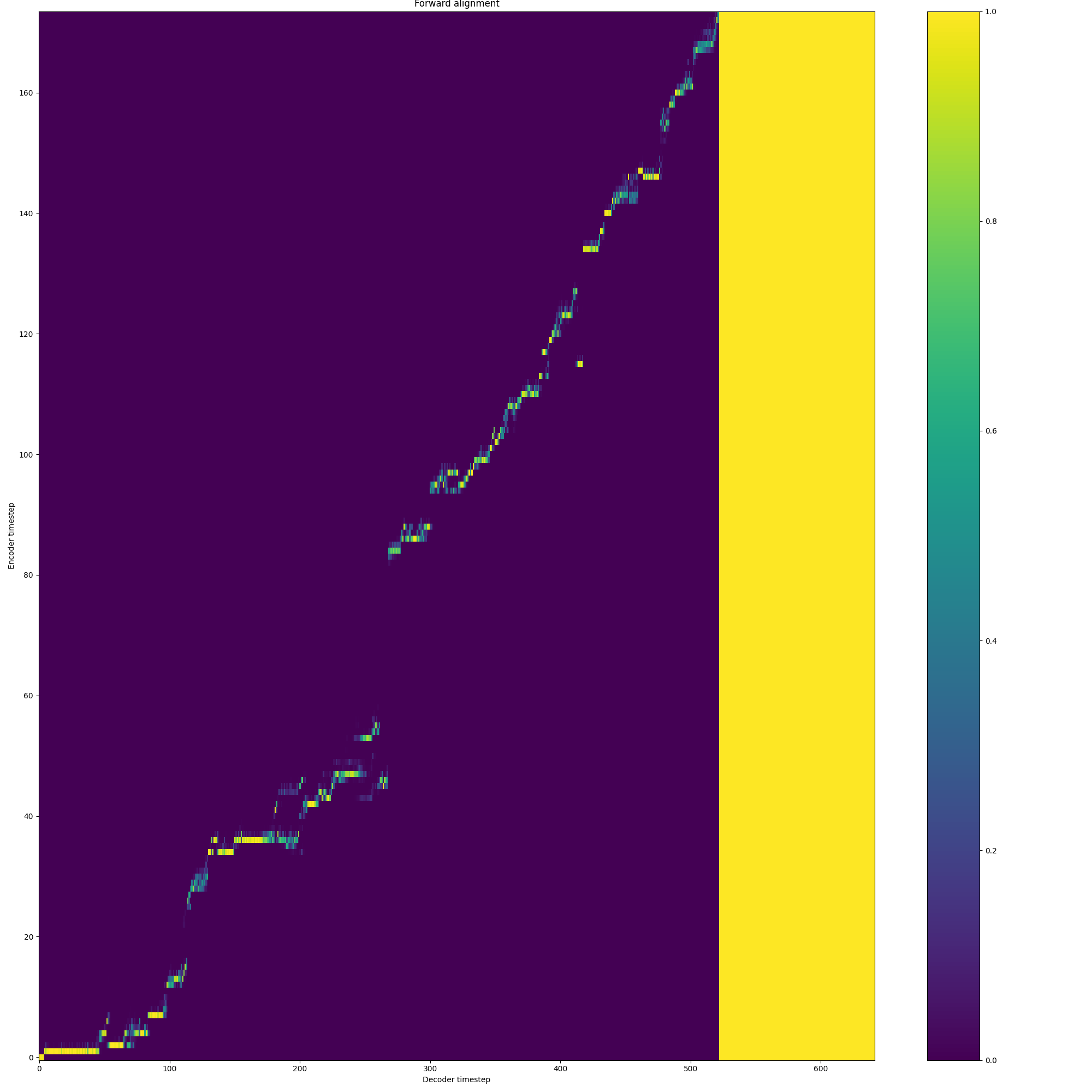}
  \caption[OverFlow Alignment]{Example of OverFlow alignment, displaying irregular patterns that might indicate alignment issues.}
  \label{fig:overflow-alignment}
\end{minipage}%
\end{figure}

\begin{figure}[ht]
\centering
\includegraphics[width=0.6\textwidth]{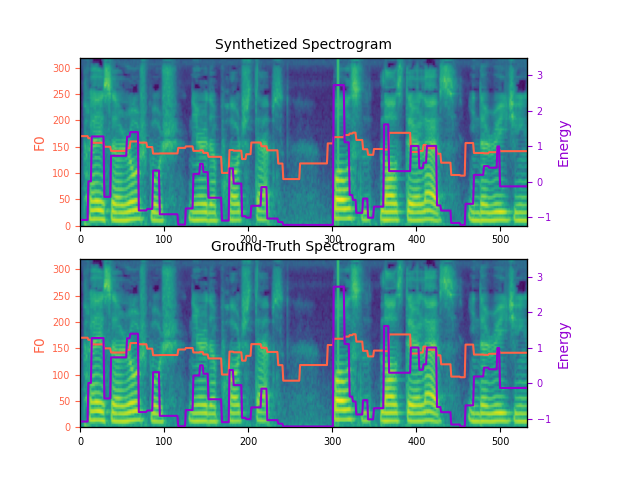}
\caption[FastSpeech2 Mel-Spectrogram Comparison]{Comparison of mel-spectrograms in FastSpeech2, highlighting the prediction of mel-spectrogram, f0, and energy.}
\label{fig:fastspeech2_mel-spectrogram}
\end{figure}

\subsection{Analysis of Alignment and Mel-Spectrogram Graphs}
By scrutinizing alignment charts, one can glean intuitive insights about the model's attention and alignment. The following are key indicators that can guide researchers in evaluating a model based on its alignment chart:

Clear Diagonal Lines: A quintessential alignment chart predominantly exhibits a distinct diagonal line, signifying a consistent alignment between input and output sequences. This indicates that the model is sequentially processing the input and generating the output in the anticipated manner.
Disjointed or Scrambled Lines: Instances where the alignment chart manifests disjointed, scrambled, or any other irregular patterns signal potential alignment challenges. Such discrepancies could arise from inconsistencies in training data, parts of the model not learning aptly, among other issues.
Diving deeper into specific models, FastPitch is a Transformer-based text-to-speech synthesis model that diverges from the standard Transformer architecture by incorporating unique features to enhance its performance and efficiency. Within FastPitch, two distinct alignment charts can be discerned:

Alignment: Representing the traditional alignment chart, it delineates the relationship between input text (like words or phonemes) and the generated acoustic features (e.g., mel-spectrogram). This alignment is a direct output of FastPitch's attention mechanism, indicating how the model aligns the input text during the generation of each acoustic feature segment.
Alignment Head: With multiple attention heads, FastPitch learns diverse attention patterns. One specific head, known as the “alignment head”, is tailored to learn the alignment dynamics between the text and acoustic features. Its specialized design often results in more explicit and coherent outputs, offering users a clearer alignment perspective.

From the various alignment graphs of different models, a multifaceted analysis and prediction can be deduced. In the graph of Tacotron2, there are noticeable discontinuities. Such interruptions might lead to moments of ambiguity during the text-to-speech conversion process, thereby affecting the fluency of the sound. Compared to the original FastPitch Alignment, the FastPitch with head showcases a more accurate alignment, indicating superior performance. FastPitch with multi-head attention should produce the best results among selected models.

Upon examining the Glow-TTS graph, it becomes evident that the lines do not fully reach the diagonal. This suggests potential alignment issues when the model processes certain data, possibly resulting in a suboptimal voice output. The alignment graph of OverFlow is particularly unsatisfactory. Such alignment discrepancies could lead to discontinuous and muffled sound, severely compromising the final synthesized voice quality. 

The alignment patterns observed in VITS are notably coherent and consistent, suggesting a strong correlation between the input and output sequences. Given this efficient alignment, it's anticipated that VITS would deliver high-quality and accurate results in its speech synthesis tasks.

In contrast to the aforementioned models, FastSpeech2 demonstrates formidable capability in spectrogram synthesis. The generated Mel-spectrogram aligns closely with the ground-truth spectrogram. Moreover, the comparison of f0 and Energy also reveals high consistency. This high degree of alignment suggests that FastSpeech2 is likely to produce excellent synthesized voice outcomes.

SpeechT5, akin to other pure Transformer-based models, does not produce an explicit, interpretable alignment chart. This stems from their reliance on the inherent attention mechanism to implicitly align input-output, as opposed to explicit alignments. The self-attention mechanism in these models captures and encodes alignment relations within hidden states, thereby not necessitating a distinct alignment chart.

\subsection{Prediction from the Graphs}

Regarding the SpeechT5 model, direct assessments based on alignment visualizations prove challenging, as it doesn't generate interpretable alignment charts. However, when comparing the models for which we do have visual insights, our analysis leads us to the following qualitative ranking in terms of expected performance: FastPitch appears to be at the forefront, followed closely by VITS. FastSpeech2 then comes next, displaying strong correspondence, but not as pronounced as the previous two. Tacotron2 seems to show some potential misalignments, placing it behind the aforementioned models. Glow-TTS, while possessing a distinct alignment pattern, doesn't seem to align perfectly diagonally, hinting at potential synthesis challenges. Finally, OverFlow appears to have the most irregular patterns, which might indicate significant alignment issues that could affect its output quality.

\section{Model Fine-tuning Time}

Table \ref{tab:training_times} presents the training durations for various state-of-the-art speech synthesis models. These times provide insight into the computational effort required to train each model. Tacotron2, for instance, has one of the lengthier training periods, spanning almost two hours. On the other end of the spectrum, SpeechT5 exhibits a surprisingly swift training duration of just 15 minutes. Such differences can be attributed to various factors including model complexity, dataset size, and optimization techniques employed.

\begin{table}[h]
\centering
\begin{tabularx}{0.6\textwidth}{|X|r|}
\hline
\textbf{Model} & \textbf{Training Time} \\
\hline
Tacotron2 & 1.803 hours \\
\hline
FastSpeech2 & 51m26s \\
\hline
FastPitch & 43m31s \\
\hline
GlowTTS & 20.31 min \\
\hline
OverFlow & 32 min \\
\hline
VITS & 4 hours 25 min \\
\hline
SpeechT5 & 15 min \\
\hline
\end{tabularx}
\caption{Training time for various speech synthesis models.}
\label{tab:training_times}
\end{table}

\section{MOS Evaluation}

In the realm of Text-to-Speech (TTS) research, evaluating the quality and naturalness of the generated speech is of paramount importance. To this end, we employ the Mean Opinion Score (MOS) as our primary metric of assessment. MOS is a widely-accepted subjective evaluation method, wherein participants are tasked with rating specific speech samples based on their perceptual experience. The scoring typically spans from 1 to 5, where 1 indicates "Poor" quality, and 5 denotes "Excellent". This scoring mechanism provides us with an intuitive and quantifiable means to gauge the performance of our TTS system. By aggregating a series of MOS scores, we can attain a comprehensive, objective feedback on the produced speech, thus aiding in further model and technique refinement. However, it's essential to note that while MOS is a potent tool, it is still anchored in human subjectivity, which may vary based on listeners' backgrounds, experiences, and expectations \cite{MOS}.

In the evaluation phase of my text-to-speech model, I used each fine-tuned model to generate a 5-second audio clip for testing with the phrase: 'Here is Ze, and this is my honour project testing'." I conducted a subjective listening test involving 38 participants who were familiar with my voice. The primary objective of this test was to ascertain the verisimilitude of the generated voice to my own. The participants were tasked with rating the samples across four key dimensions:

\begin{enumerate}
    \item Timbral Similarity:
    Participants assessed the degree of resemblance between the synthesized voice and my natural voice. This dimension gauges how well the TTS system has been able to capture and replicate the unique characteristics inherent to my vocal identity.
    \item Naturalness:
    This metric evaluates the lifelikeness of the synthesized speech. Here, listeners determined whether the output sounded authentically human or exhibited artifacts typical of machine-generated voices.
    \item Fluency: 
    Under this criterion, the smoothness and continuity of the speech were examined. Participants listened for any stutters, hesitations, or unnatural breaks that might disrupt the flow, making the speech sound less fluid.
    \item Pitch and Emotion: 
    This metric is concerned with the tonal quality and emotional expressiveness of the speech. Participants judged if the voice conveyed varied emotions and intonations or if it sounded monotonous, akin to the flat tonality often associated with robotic voices.
\end{enumerate}

These subjective evaluations provided valuable insights into the strengths and potential areas of improvement for my text-to-speech system, ensuring its ability to produce outputs that are both high-fidelity and emotionally resonant.

\begin{table}[h!]
\centering
\begin{tabular}{|c|c|c|c|c|}
\hline
\textbf{Model} & \textbf{Timbral Similarity} & \textbf{Naturalness} & \textbf{Fluency} & \textbf{Pitch and Emotion} \\
\hline
\textbf{Tacotron2} & 4.605 ± 0.85 & 4.395 ± 0.089 & 3.211 ± 0.106 & 3.605 ± 0.100 \\

\textbf{VITS} & 4.447 ± 0.081 & 4.289 ± 0.108 & 3.263 ± 0.129 & 3.263 ± 0.098 \\

\textbf{Glow-TTS} & 2.579 ± 0.123 & 2.026 ± 0.026 & 3.237 ± 0.070 & 2.842 ± 0.144 \\

\textbf{FastSpeech2} & 2.000 ± 0.084 & 1.579 ± 0.081 & 1.816 ± 0.091 & 1.711 ± 0.106 \\

\textbf{OverFlow} & 1.132 ± 0.056 & 3.342 ± 0.078 & 4.237 ± 0.070 & 2.605 ± 0.080 \\

\textbf{FastPitch} & 4.263 ± 0.072 & 4.474 ± 0.082 & 3.947 ± 0.075 & 3.789 ± 0.108 \\

\textbf{SpeechT5} & 4.132 ± 0.120 & 4.289 ± 0.075 & 4.526 ± 0.118 & 4.447 ± 0.105 \\
\hline
\end{tabular}
\caption{MOS score of various Text-to-Speech models across key metrics: Timbral Similarity, Naturalness, Fluency, and Pitch \& Emotion}
\label{tab:MOS}
\end{table}

The table showcases the performance of various text-to-speech models across four metrics: Timbral Similarity, Naturalness, Fluency, and Pitch \& Emotion. Each model's scores were measured relative to the familiarity of the speaker's voice, with higher scores indicating better performance.

Timbral Similarity: This metric gauges the degree of resemblance between the synthesized voice and the original speaker's voice. Tacotron2 and VITS exhibit superior performance in this area, scoring 4.605 and 4.447, respectively. This indicates that both models are adept at capturing the specific characteristics of the original voice. On the contrary, OverFlow seems to struggle in this metric with a score of 1.132.

Naturalness: Referring to how lifelike and real the synthesized voice sounds, the highest performance is noted from FastPitch at 4.474. This suggests that its generated voice is almost indistinguishable from a real human voice. In contrast, FastSpeech2 possesses a score of 1.579, implying that its output might sound more mechanical or artificial.

Fluency: This metric evaluates the smoothness of the synthesized speech, ensuring there are no stutters or breaks. SpeechT5 takes the lead in this domain, securing a score of 4.526. This highlights its proficiency in delivering speech that flows seamlessly. Conversely, FastSpeech2 trails behind other models with a score of 1.816.

Pitch and Emotion: A critical factor, as it pertains to the emotional expressiveness and tonal variations in the voice. SpeechT5 again leads the pack with 4.447, meaning its output contains varied and human-like emotional tones. In contrast, FastSpeech2 and OverFlow score lower, hinting at potential monotony or lack of emotional depth in their outputs.

An overall score can be calculated as an average across all four metrics for each model. 

\begin{table}[h]
\centering
\caption{Average Scores of Various Text-to-Speech Models}
\begin{tabular}{|c|c|}
\hline
\textbf{Model} & \textbf{Average Score} \\
\hline
Tacotron2 & 3.954 \\
\hline
VITS & 3.816 \\
\hline
Glow-TTS & 2.671 \\
\hline
FastSpeech2 & 1.777 \\
\hline
OverFlow & 2.829 \\
\hline
FastPitch & 4.118 \\
\hline
SpeechT5 & 4.349 \\
\hline
\end{tabular}
\end{table}

Based on those scores, SpeechT5 has the highest overall score of 4.349, making it the top-performing model in this comparison.

FastSpeech2 holds the lowest average score of 1.777, suggesting it might be the least effective in replicating voice characteristics as per the chosen metrics.

Remember, these averages provide a summarized perspective and may not account for nuances in specific use-cases.

%% file: Discussion.tex
\chapter{Discussion}
\section{Interpreting the Findings}

Based on the feedback and synthesized audio results I received, OverFlow was unable to capture the timbre of my voice. Considering that the pre-trained model was based on LJSpeech, the generated voice remained feminine. However, the pauses and intonation did bear some resemblance to mine, suggesting that its capability for transfer learning with a small dataset is lacking. In contrast, the VITS model supports seamless interchange among Chinese, English, and Japanese languages. The ability to input in any of these three languages and obtain outputs in all three is truly remarkable.

SpeechT5 can handle a variety of spoken language tasks, including automatic speech recognition, speech synthesis, speech translation, voice conversion, speech enhancement, and speaker identification, as cited from \cite{ao2021speecht5}. Furthermore, the fact that it requires only a mere 15 minutes of training to achieve this is astonishing. The significance of Tacotron2's work cannot be understated, as many subsequent models were inspired and improved upon its foundational ideas.

Upon evaluation of the selected Text-to-Speech models, it became evident that the application of transfer learning in a few-shot, low-resource setting varied significantly across models. While some models exhibited a commendable ability to harness pre-existing knowledge and apply it effectively to the limited dataset, others struggled to maintain their previously acclaimed high performance.

\section{Hypotheses Verification}

\subsection{Effectiveness of Transfer Learning}
Our first hypothesis posited that transfer learning significantly enhances the capability of TTS models to perform effectively on smaller datasets. The experiments largely verified this, with a marked improvement in voice quality, fluency, and naturalness in models leveraging transfer learning compared to those trained from scratch on the small dataset.

\subsection{Optimal TTS Model Identification}
Our second hypothesis suggested the existence of an optimal TTS model that offers a harmonious blend of rapid training, minimal data dependence, and exemplary voice quality. From our experiments, SpeechT5 emerged as a potential contender for this title, showcasing a promising balance of the stated criteria.
\section{Comparison with Prior Work}

Comparing our findings with previous research, we observed a consistent trend. Earlier studies, such as those by \cite{transfer_learning_1} and \cite{transfer_learning_2}, also highlighted the challenges and potential of transfer learning in TTS. However, our study offers a more extensive comparison, especially in the context of the few-shot, low-resource scenario.

\section{Limitations}

While our research provides crucial insights, it comes with its set of limitations. One of the limitations of this study is that cannot guarantee that every parameter chosen is optimal for each individual model. The choice of the dataset, though customized, might not represent all types of low-resource scenarios. Additionally, the models evaluated were bound by our computational constraints, and there could be newer models in the horizon that offer even better transfer learning capabilities.

\section{Ethical Considerations}

The rise of voice synthesis technology has inadvertently given a boost to the nefarious activities of scammers. Recently, there has been a surge in incidents where fraudsters clone the voices of family members or friends to deceive unsuspecting individuals. With advancements in technology, it is now possible to replicate a person's voice with just a few samples of their speech. Additionally, the digital realm is witnessing the emergence of AI-driven artists, such as the famed "Fake Drake". A song titled "Heart on my Sleeve \footnote{https://www.youtube.com/watch?v=7HZ2ie2ErFI}" featuring both AI Drake and the singer The Weeknd gained immense popularity online, racking up a staggering 4.3 million plays. This collaboration between artificial and human voices has sparked widespread interest and debate among netizens.

Text-to-Speech (TTS) technologies have several ethical considerations that change as the technology advances. As TTS systems improve, they can accurately mimic a person's voice with a few samples. This feature raises worries about cloned voices being used for identity theft or misinformation operations.

Who owns a voice? Have they given explicit permission if a person's voice is cloned and used in media, advertisements, or other commercial activities? Informed permission for voice cloning is essential to prevent legal and ethical issues. TTS can make "deepfake" audio recordings of people saying things they never did. Fake recordings may defame, manipulate public opinion, and even influence elections.

As AI-generated voices grow more common, voice-over artists, narrators, and vocalists may struggle. The economic effects of replacing human talent with AI voices must be considered.
In music and entertainment, AI-driven musicians like "Fake Drake" raise doubts about authenticity. Can an AI-generated song ever be as emotive as a human-written and performed one? TTS's capacity to alter speech tones might be used to emotionally manipulate listeners. In advertising, an AI may change its pitch or tone to evoke trust or urgency, posing ethical questions about autonomy and manipulation.

Should listeners know the difference between TTS and human voices? News reporting and education need transparency because trust and authenticity matter.
TTS technologies have the potential to disrupt many industries, but they also bring ethical issues. As with many innovative technology, safe usage requires continuing discourse, strict laws, and public awareness efforts.